\documentclass{aa}
\usepackage[varg]{txfonts}

\usepackage{graphicx}
\usepackage{tabularx}

\usepackage{natbib,twoopt}
\usepackage[breaklinks=true]{hyperref} 
\hypersetup{
  colorlinks=true,   
  urlcolor=blue,     
  linkcolor=blue     
}

\bibpunct{(}{)}{;}{a}{}{,}             
\makeatletter
  \newcommandtwoopt{\citeads}[3][][]{\href{http://adsabs.harvard.edu/abs/#3}%
    {\def\hyper@linkstart##1##2{}%
     \let\hyper@linkend\@empty\citealp[#1][#2]{#3}}}
  \newcommandtwoopt{\citepads}[3][][]{\href{http://adsabs.harvard.edu/abs/#3}%
    {\def\hyper@linkstart##1##2{}%
     \let\hyper@linkend\@empty\citep[#1][#2]{#3}}}
  \newcommandtwoopt{\citetads}[3][][]{\href{http://adsabs.harvard.edu/abs/#3}%
    {\def\hyper@linkstart##1##2{}%
     \let\hyper@linkend\@empty\citet[#1][#2]{#3}}}
  \newcommandtwoopt{\citeyearads}[3][][]%
    {\href{http://adsabs.harvard.edu/abs/#3}
    {\def\hyper@linkstart##1##2{}%
     \let\hyper@linkend\@empty\citeyear[#1][#2]{#3}}}
 \renewcommand*\aa@pageof{, page \thepage{} of \pageref*{LastPage}} 
\makeatother

\usepackage{booktabs}
\usepackage{epstopdf}

\usepackage[capitalise]{cleveref}

\usepackage{bm}
\usepackage{physics}

\allowdisplaybreaks

\begin{document}

\title{A spectroscopic investigation of thermal instability for cylindrical equilibria with background flow}

\titlerunning{Thermal instability and helical flow}

\author{J. Hermans\inst{1}\and R. Keppens\inst{1}}

\institute{$^1$ Centre for mathematical Plasma-Astrophysics, Celestijnenlaan 200B, 3001 Leuven, KU Leuven, Belgium} 

\date{Received date /
Accepted date }
 
\abstract{Flows are omnipresent and govern the dynamics of plasma. Solar tornadoes are a class of apparently rotating
prominences, that might be formed by thermal instability. In spectroscopic studies on thermal instability background flow is commonly neglected.}{We here determine the effect of background flow on thermal instability in cylindrical magnetic field configurations. How various parameters affect the distribution of eigenmodes in the MHD spectrum is also explored. We investigate whether discrete thermal modes exist.}{In an analytical study, we extend upon the literature by including a generic background flow in a cylindrical coordinate system. The non-adiabatic MHD equations are linearised, Fourier-analysed, and are examined to understand how a background flow changes the continua. An approximate expression for discrete thermal modes is derived using a WKB analysis. The analytical results are then verified for a benchmark equilibrium using the eigenvalue code \textit{Legolas}. The eigenfunctions of discrete thermal modes are visualised in 2D and 3D.}{The thermal continuum is Doppler-shifted due to the background flow, just like the slow and Alfv\'en continua. Discrete modes are altered because the governing equations contain flow-related terms. An approximate expression to predict the appearance of discrete thermal modes based on the equilibrium parameters is derived. All analytical expressions match the numerical results. The distribution of the density perturbations of the discrete thermal modes is not a uniform or singular condensation, due to the shape of the eigenfunctions and the dependence of the assumed waveform on the coordinates and wavenumbers. 3D visualisation of the total velocity field shows that the helical field is heavily influenced by the radial velocity perturbation.}{We derived analytic expressions for non-adiabatic MHD modes of a cylindrical equilibrium with background flow and verified them using a coronal equilibrium. However, the equations are valid for and can be applied in other astrophysical environments.}


\maketitle

\graphicspath{{Figures/}}

\section{Introduction}\label{Sec: Intro_TIflow}

The solar atmosphere is a very dynamic environment. Reconnection is the cause of energetic phenomena such as solar flares and jets. Jets may be represented as fast, rotating outflows \citepads{2021RSPSA.47700217S}. This transient flow of plasma has been observed all around the solar corona \citepads{2021A&A...656L..13C,2023ApJ...944...19L}. Waves and quasi-periodic flows are also common \citepads{2015SoPh..290..399D,2021SSRv..217...76B}. Upflows from the transition layer and corona are considered the driving force behind the slow solar wind \citepads{2021A&A...651A.112B,2023A&A...673A..74B}. Furthermore, prominences are involved in violent eruptions, creating shock waves throughout the corona and launching coronal mass ejections, CMEs, into space \citepads{2015ASSL..415..411W}. Prominences are host to a variety of flows \citepads{2015ASSL..415...79K}. 

Solar tornadoes are a kind of prominences with plasma that appears to be rotating around a vertical axis \citepads{1932ApJ....76....9P}. The phenomenon has recently been the topic of the review paper by \citetads{2023SSRv..219...33G}. Many observations were made last decade, see e.g. \citetads{2012ApJ...756L..41S,2012ApJ...752L..22L,2013ApJ...774..123W,2014ApJ...785L...2S,2018ApJ...852...79Y}. Line-of-sight Doppler velocities were measured by \citetads{2012ApJ...761L..25O}. The red- and blue-shifted pattern was interpreted as rotation around a vertical axis. \citetads{2018ApJ...852...79Y} observed two tornadoes with both coherent, stable red and blue Doppler shifts in favour of this interpretation. Models of helical magnetic structures with cool plasma flowing helically are constructed by \citetads{2015ApJ...808L..23L} and \citetads{2018PhPl...25e4503O}. For the tornado studied by \citetads{2012ApJ...752L..22L}, the rotational axis is believed to be parallel to the axis of a horizontal flux rope. Such rotational flow in a prominence cavity has recently been studied in a multidimensional simulation of a flux rope prominence \citepads{2023ApJ...953L..13L}. Prominences are dense structures that are assumed to be suspended in the corona by the magnetic field. \citetads{2015ApJ...808L..23L} showed that prominence mass can be supported by a vertical helical field if the magnetic field is highly twisted and/or significant poloidal upflows are present. As the rotating nature of tornado prominences is highly controversial, counterarguments and interpretations are provided \citepads{2023SSRv..219...33G}. \citetads{2014SoPh..289..603P} argued that counterstreaming flows \citepads{1991A&A...252..353S} and oscillations on horizontal threads cause the illusion of rotation. The projection of the 3D structure of a horizontal prominence onto the 2D plane of view can make the field lines look elliptical \citepads{Gunar_2018}.  

Local thermal instability is the preferred method to form cool condensations in hot media, such as prominences in the solar corona. \citetads{1953ApJ...117..431P} noticed that the increase of radiative losses with decreasing temperature can lead to runaway cooling. The theory of thermal instability was first described by \citetads{1965ApJ...142..531F}. \citetads{2019A&A...624A..96C}, \citetads{2020A&A...636A.112C}, and \citetads{2021A&A...655A..36H} performed multidimensional simulations of condensation formation by the thermal mode in a coronal volume.

The spectral study of the MHD waves and instabilities is a fundamental methodology to understand plasmas and their evolution. Frequencies of the eigenmodes can be represented as a spectrum on the complex frequency plane. Depending on the chosen wave-form, the modes with a positive imaginary part of the frequency are unstable and lead to instabilities. The general structure of the eigenvalue spectrum of a cylindrical equilibrium consists out of two continua \citepads{1974PhFl...17.1471A}, the Alfv\'en and the slow continua, due to inhomogeneity and radial dependence of the background variables. Discrete modes can arise that cluster towards the edges of the continua \citepads{1974PhFl...17..908G,1984PhyD...12..107G}. Due to the complexity it is challenging to interpret the complete MHD spectrum of waves and instabilities for equilibria with both axial and azimuthal flow. The first systematic studies of the spectrum of cylindrical plasmas with flow were performed by \citetads{1981JMP....22.2080H,1983PhFl...26..230H}. \citetads{1987PhFl...30.2167B} examined the MHD spectrum for cylindrical plasmas with axial flow and obtained local criteria for clustering of discrete modes, whereas \citetads{2004JPlPh..70..651W} included both azimuthal and axial flow. The effect of rotational flows on magnetoacoustic surface and body modes of an adiabatic photospheric magnetic flux tube was recently studied by \citetads{2023MNRAS.518.6355S}. The existence of the thermal continuum, analogous to the Alfv\'en and slow continua, was demonstrated by \citetads{1991SoPh..134..247V}. This arises for an equilibrium with radial variation when non-adiabatic terms are included. They also derived approximate expressions to predict the occurance of discrete thermal modes using a WKB analysis, similar to \citetads{1984PhyD...12..107G}. However, background flow is typically neglected in spectroscopic studies of thermal instability.

In recent years the numerical study of MHD spectroscopy has been revived by the development of the eigenvalue code \textit{Legolas} \citepads{2020ApJS..251...25C,2022JPlPh..88c9021D,2023arXiv230710145C}. It is an open-source code to solve the eigenvalue problem of the linearised and Fourier-analysed MHD equations. It provides the eigenfunctions and the corresponding eigenfrequencies, which can be represented as a spectrum on the complex plane. \citetads{2021SoPh..296..143C} used \textit{Legolas} to study the eigenmodes of a model solar atmosphere, focusing on the thermal modes. A large scale, systematic investigation of the scaling of the tearing growth rate with resistivity, density, and velocity parameters was recently completed using \textit{Legolas} \citepads{alma9993464882001488}.

In \cref{Sec: TI_flow_ana} we extend upon the work of \citetads{1991SoPh..134..247V} by including a generic background flow in a cylindrical coordinate system. The non-adiabatic MHD equations are linearised, Fourier-analysed, and transformed into two coupled first-order differential equations. These are examined to understand how a background flow changes the continua. The numerical investigation is presented in \cref{Sec: num_inv_cont}. We setup a benchmark equilibrium and explore its eigenvalue spectrum using \textit{Legolas}. The benchmark has a force-free Gold-Hoyle magnetic field and helical flow field. The derived analytic expressions of the previous section are verified. Several equilibrium parameters are varied to study the effect on the eigenmodes. In \cref{Sec: discretemodes} discrete thermal modes are the topic of interest. An approximate expression is derived using a WKB analysis. This expression can be used to predict the appearance of discrete thermal modes based on the equilibrium parameters. The modes are then investigated numerically and the previously obtained expression is tested. Lastly, the eigenfunctions of the discrete eigenmodes are visualised. We present a summary of the results, discussion, and conclusions in \cref{Sec: sum-dis_TIflow}.

\section{Analytic investigation}\label{Sec: TI_flow_ana}

In this section we derive the expression for the thermal continuum following closely the work by \citetads{1991SoPh..134..247V}. The additional physical effect that we take into account is a background flow. The solar tornadoes that we are interested in are vertical column-like structures. A cylindrical coordinate system is thus the most convenient way to describe the plasma. In order to study thermal instability one needs to include radiative losses. We also include a background heating to match the losses and obtain a thermal equilibrium to investigate. Since thermal conduction is a relevant and omnipresent effect in the solar corona, it is included. The influence of thermal conduction on thermal instability has been shown in many works \citepads{1965ApJ...142..531F,1991SoPh..134..247V,2010ApJ...720..652S}. For simplicity we only include parallel conduction. The inclusion of perpendicular conduction can modify the thermal modes drastically \citepads{1991SoPh..134..247V}. In fact, perpendicular conduction replaces the continuum with a closely packed set of quasi-continuum modes that are in essence discrete. The eigenfunctions of those quasi-continuum modes might have a localised, rapidly spatially-varying stucture, providing us with a natural explanation for finestructure in solar prominences. Since the thermal continuum does not strictly exist with perpendicular conduction included, we do not include it here. We also do not include gravity, as it is not in the radial direction when considering a vertical cylinder.

To start, we consider a fully ionised plasma under influence of aforementioned non-adiabatic effects. The following set of dimensionless MHD equations are appropriate to describe this medium:
\begin{align}\label{Eq: MHDeq_TIflow}
    \pdv{\rho}{t} = &- \nabla \cdot (\bm{v}\rho),\\
    \rho \pdv{\bm{v}}{t}  = &-\nabla p - \rho \bm{v} \cdot \nabla\bm{v} + (\nabla\crossproduct\bm{B} )\crossproduct \bm{B},\\
    \begin{split}\label{Eq: MHDeq_TIflow_last}
    \rho\pdv{T}{t} =  &-\rho \bm{v}\cdot \nabla T - (\gamma-1)p \nabla\cdot\bm{v} - (\gamma-1)\rho \mathcal{L}\\
    & + (\gamma-1)\nabla \cdot (\bm{\kappa} \cdot \nabla T), 
    \end{split}\\
    \pdv{\bm{B}}{t} = & - \bm{B} (\nabla\cdot\bm{v}) + (\bm{B} \cdot \nabla)\bm{v} - (\bm{v} \cdot \nabla)\bm{B},
\end{align}

\noindent with the magnetic field in units, where $\mu_0 = 1$. The quantities $\rho$, $T$, $p$, $\bm{v}$, and $\bm{B}$ are the density, temperature, pressure, velocity, and magnetic field, respectively. The function $\mathcal{L}$ is the net heat-loss function and is defined as
\begin{equation}
    \mathcal{L} = \rho \Lambda(T) - \mathcal{H}.
\end{equation}

It represents the difference in energy loss between the radiative cooling, for which we assume an optically thin plasma, and an unknown heating function $\mathcal{H}$. The optically thin cooling function or curve,  $\Lambda(T)$, describes radiative energy loss in function of the temperature. A wide variety is used in the literature. The effect of the cooling curve on numerical simulations of thermal instability is discussed in \citetads{2021A&A...655A..36H}.

The last term of the right-hand side of the energy equation is related to thermal conduction. $\bm{\kappa}$ is the thermal conduction tensor. In magnetised plasmas thermal conduction is anisotropic and typically respresented as
\begin{equation}\label{Eq: conductionTIflow}
    \bm{\kappa} = \kappa_\parallel \bm{e}_B\bm{e}_B + \kappa_\perp (\bm{I} - \bm{e}_B\bm{e}_B),
\end{equation}

\noindent where $\bm{I}$ denotes the unit tensor and $\bm{e}_B = \bm{B}/B$ is the unit vector along the magnetic field lines. The coefficients $\kappa_\parallel$ and $\kappa_\perp$ denote the conductivity coefficients parallel and perpendicular to the local direction of the magnetic field, respectively. Explicit formulae in terms of temperature, density, and magnetic field strength for astrophysical applications can be found in \citetads{1962pfig.book.....S}. For the solar corona the perpendicular coefficient is typically twelve orders of magnitude smaller than the parallel coefficient \citepads{1965RvPP....1..205B}. Hence, as in most works we therefore neglect perpendicular thermal conduction, i.e. take $\kappa_\perp = 0$. This also simplifies the analytic derivations. 

We consider an equilibrium around which we linearise the equations. As our main interest is helical flows and magnetic fields, working in a cylindrical coordinate system is most beneficial. We use the convential meaning of the variables, ($r$, $\theta$, $z$) being the radial coordinate, azimuthal angle and axial coordinate, respectively. We consider an axisymmetric tube with all background parameters varying in only one dimension, the radial direction. The background quantities are denoted with a subscript `0'.  
The ideal gas law defines the relation between the background pressure, density and, temperature as 
\begin{equation}
    p_0(r) = \rho_0(r) T_0(r)
\end{equation}

\noindent in the dimensionless notation adopted in this work.

We do include a background flow. This flow is stationary, i.e. not varying in time. The background profiles of the equilibrium flow are given by 
\begin{equation}
    \bm{v}_0 = v_{0\theta}(r) \bm{e}_{\theta} +   v_{0z}(r) \bm{e}_{z}.
\end{equation}

\noindent The background magnetic field is given by 
\begin{equation}
     \bm{B}_0 = B_{0\theta}(r) \bm{e}_{\theta} +   B_{0z}(r) \bm{e}_{z}.
\end{equation}

\noindent The divergence-free condition of the magnetic field, $\nabla \cdot \bm{B} = 0$, is automatically satisfied for our background magnetic field because the non-radial components only depend on the radial coordinate and $B_r = 0$. Similarly, the equilibrium flow is incompressible, but compressibility effects are fully accounted for in the linearised equations.

Furthermore, the background profiles are at equilibrium, meaning that they have to satisfy the mechanical and thermal equilibrium. In order to obtain mechanical equilibrium the following radial force balance needs to be satisfied for the equilibrium profiles.
\begin{equation}\label{Eq: Mechequi}
    D\left( p_0 + \frac{1}{2}B_0^2\right) = -\frac{B_{0\theta}^2}{r} + \frac{\rho_0 v_{0\theta}^2}{r}
\end{equation}
\noindent We denote the derivative in the direction of $r$, i.e. $\pdv{}{r}$, as $D$. Hence, we follow the notation of \citetads{1991SoPh..134..247V}. Compared to static equilibria, such as in the case of \citetads{1991SoPh..134..247V}, the background flow introduces an additional term on the right-hand-side. This term corresponds to the centrifugal force and is only apparent in a cylindrical configuration. Equilibria with rotational flow and static equilibria are thus inherently slightly different. The axial flow profile $v_{0z}$ is of no relevance to the mechanical equilibrium and can be chosen freely. 

Since we are interested in thermal instabilities, we require an initial state in thermal equilibrium. Under the assumption of neglected perpendicular conduction, this can be written as
\begin{equation}
    \rho_0 \mathcal{L}_0 = 0, 
\end{equation}
\noindent where $\mathcal{L}_0$ is the heat-loss function at the initial state. Intuitively this condition means that the energy-loss by the optically thin radiative cooling needs to be balanced by the background heating at $t=0$. As an equation this means
\begin{equation}
    \mathcal{H}_0 = \rho_0 \Lambda(T_0).
\end{equation}

We consider the Eulerian perturbations to be small, so we can expand around the equilibrium in the form 
\begin{equation}
    f(\bm{r},t) = f_0(r) + f_1(\bm{r},t),
\end{equation}

\noindent where $f$ represents any of the physical quantities and the subscript `1' is used for the dynamic, i.e. perturbed, part. The equilibrium is constant in time, such that temporal derivatives vanish. We are only interested in linear perturbations and can ignore all higher-order terms. Substituting these expansions into the MHD equations, \crefrange{Eq: MHDeq_TIflow}{Eq: MHDeq_TIflow_last}, yields 
\begin{align}
    \pdv{\rho_1}{t} = &- \nabla \cdot (\rho_0\bm{v}_1) -  \nabla \cdot (\rho_1 \bm{v}_0),\\
    \begin{split}
    \rho_0 \pdv{\bm{v}_1}{t}  = &-\nabla p_1 - \rho_0 \bm{v}_0 \cdot \nabla\bm{v}_1 - \rho_0 \bm{v}_1 \cdot \nabla\bm{v}_0 - \rho_1 \bm{v}_0 \cdot \nabla\bm{v}_0\\
    & + (\nabla\crossproduct\bm{B}_1 )\crossproduct \bm{B}_0+ (\nabla\crossproduct\bm{B}_0 )\crossproduct \bm{B}_1,
    \end{split}\\
    \begin{split}
    \rho_0\pdv{T_1}{t} =  &-\rho_0 \bm{v}_0\cdot \nabla T_1 - \rho_0 \bm{v}_1\cdot \nabla T_0 - (\gamma-1)p_0 \nabla\cdot\bm{v}_1\\ 
       & - (\gamma-1)\rho_0 [ \rho_1 \mathcal{L}_{\rho} + T_1 \mathcal{L}_{T} ]\\ 
       & + (\gamma-1) \bm{B}_0 \cdot \nabla \left[ \frac{\kappa_{\parallel,0}}{B_0^2} \bm{B}_0 \cdot \nabla T_1 \right]\\ 
       & + (\gamma-1) \bm{B}_0 \cdot \nabla \left[ \frac{\kappa_{\parallel,0}}{B_0^2} \bm{B}_1 \cdot \nabla T_0 \right],
    \end{split}\\
    \begin{split}
    \pdv{\bm{B}_1}{t} = & ~(\bm{B}_0 \cdot \nabla)\bm{v}_1 + (\bm{B}_1 \cdot \nabla)\bm{v}_0 - \bm{B}_0 (\nabla\cdot\bm{v}_1) - (\bm{v}_0 \cdot \nabla)\bm{B}_1 \\
    & - (\bm{v}_1 \cdot \nabla)\bm{B}_0.
    \end{split}
\end{align}

\noindent We used the equilibrium conditions of mechanical and thermal equilibrium, together with the fact that the background is stationary, $\pdv{f_0}{t}=0$. 

The quantities $\mathcal{L}_{\rho}$ and $\mathcal{L}_{T}$ are the partial derivatives of the heat-loss function with respect to density and temperature, respectively. Hence, they are given by
\begin{equation}
    \mathcal{L}_{\rho} = \left( \pdv{\mathcal{L}}{\rho} \right)_T, \quad    \mathcal{L}_{T} =\left( \pdv{\mathcal{L}}{T} \right)_\rho,
\end{equation}

\noindent to be evaluated at the equilibrium values. These derivatives are important ingredients in the instability criteria of \citetads{1965ApJ...142..531F} for the thermal instability. The derivative with respect to the density is usually positive for astrophysical plasmas. This is the case for the optically thin plasma in the solar corona. The derivative with respect to the temperature is determined by the shape of the cooling curve, as the latter is a function of temperature. For most cooling curves discussed in this work the cooling curve decreases with increasing temperature around coronal values, see e.g. \citetads{2021A&A...655A..36H}. The derivative with respect to temperature is thus negative and this leads to thermal instability. Although, thermal conduction can stabilise the medium.

Neglecting perpendicular thermal conduction simplifies the linearisation of the thermal conduction term. To linearise the thermal conduction term, we made use of the definition \cref{Eq: conductionTIflow} and the vector calculus identity
\begin{equation}
    \nabla \cdot (\bm{B}_0 f) = f (\nabla \cdot \bm{B}_0) + \bm{B}_0 \cdot (\nabla f).
\end{equation}
The first term on the right-handside vanishes because of the divergence-free background magnetic field. We also used the fact $B_r = 0$, which implies that 
\begin{equation}
    \bm{B}_0 \cdot \nabla T_0 = 0,
\end{equation}

\noindent to have no terms with the perturbed parallel conduction coefficient present. Therefore, we can drop the subscript `0' of the parallel conduction coefficient.

Since the equilibrium quantities only depend on the radial coordinates, we Fourier-analyse the perturbed quantities with respect to the ignorable coordinates, being $\theta$ and $z$. One typically introduces the wavenumber $m$ and $k$ in the azimuthal and axial directions, respectively. The time dependence is assumed exponential with a complex frequency $\omega$. The perturbed quantities can then be written as plane-waves
\begin{equation}\label{Eq: Fourierdecomp}
    f_1 (\bm{r},t) = f'(r)e^{i(m\theta + k z-\omega t)},
\end{equation}

\noindent with the primed $f'(r)$ being the amplitudes of the waves. Note that in the work by \citetads{1991SoPh..134..247V} the notation of the time dependence and frequency are slightly different. They use the exponential as $e^{st}$, hence for direct comparison one should use the substitution $s = -i\omega$ or $\omega = is$. This has as most important effect that the real and complex axis are swapped. Purely imaginary modes in the notation used in this work, such as the thermal modes, are denoted as purely real modes in Van der Linden's work.

We introduce an additional quantity, the total pressure perturbation as
\begin{equation} 
    Y = p' + B_{0\theta}B_{\theta}' + B_{0z}B_{z}',
\end{equation}

\noindent which allows us to reduce the set of eight linear MHD equations to two first order differential equations. Furthermore, we use the linearised ideal gas law to define the pressure perturbation.
\begin{equation}
    \frac{p'}{p_0} = \frac{\rho'}{\rho_0} + \frac{T'}{T_0}.
\end{equation}

Since equilibrium values are always denoted with a subscript `0', we can drop the prime on the perturbed quantities. The Fourier-analysed MHD equations are given by
\begin{align}\label{Eq: MHDfourierfirst}
    [\omega - \Omega_0]\rho =& - \frac{i}{r} D(r\rho_0 v_r) + \rho_0\left[\frac{m}{r}v_{\theta} + k v_z\right],\\
    \begin{split}
    [\omega - \Omega_0]\rho_0 v_r  =&  -i D(Y) - F B_r - \frac{2iB_{0\theta}}{r}B_{\theta} + \frac{iv^{2}_{0\theta}}{r}\rho \\
    &+\frac{2i\rho_0v_{0\theta}}{r} v_{\theta},
    \end{split}\\
    [\omega - \Omega_0]\rho_0 r v_\theta =& ~mY -rF B_{\theta} + iD(rB_{0\theta}) B_r - i\rho_0D(rv_{0\theta})  v_r,\\
    [\omega - \Omega_0]\rho_0 v_z = &~ kY - FB_{z} + iD(B_{0z}) B_r - i\rho_0 D(v_{0z}) v_r,\\
    \begin{split}
        \Biggl[ [\omega - \Omega_0] \rho_0 + i&(\gamma - 1)\left[\frac{\kappa_\parallel}{B_0^2}F^2 + \rho_0 \mathcal{L}_T\right]\Biggr] T \\
    = &-i\rho_0 D(T_0) v_r - i(\gamma - 1)\frac{p_0}{r}D(rv_r)\\
    & + (\gamma - 1)p_0 \left[\frac{m}{r} v_{\theta} + k v_z\right] -i (\gamma - 1) \rho_0 \mathcal{L}_{\rho}\rho\\
     &- (\gamma - 1) \frac{\kappa_\parallel}{B_0^2}F D(T_0) B_r,
    \end{split}\\
    [\omega - \Omega_0]B_r =& -F v_r,\\
    [\omega - \Omega_0] B_{\theta} =&  -iD(B_{0\theta} v_r) - k [B_{0z} v_{\theta} - B_{0\theta} v_z] + ir D\left(\frac{v_{0\theta}}{r}\right)B_r ,\\
    [\omega - \Omega_0] B_{z} =&  -\frac{i}{r}D(rB_{0z} v_r) + \frac{m}{r} [B_{0z} v_{\theta} - B_{0\theta} v_z] + i D(v_{0z})B_r,\\
    \begin{split}\label{Eq: MHDfourierlast}
    [\omega - \Omega_0] rY =&  ~[\omega - \Omega_0]\frac{ rp_0}{\rho_0}\rho + [\omega - \Omega_0]\frac{r p_0}{T_0} T\\
     & - irB_{0\theta} D(B_{0\theta}v_r) -iB_{0z}D(rB_{0z}v_r)\\ 
     & + r\left[\frac{m}{r}B_{0z} - kB_{0\theta}\right][B_{0z}v_{\theta} - B_{0\theta}v_z]\\
     & + \left[rB_{0\theta}D\left(\frac{v_{0\theta}}{r}\right) +B_{0z} D(v_{0z})\right]irB_r,
    \end{split}
\end{align}

\noindent where 
\begin{equation}\label{Eq: Fandds}
    F = \bm{k}\cdot\bm{B}_0 \quad \text{and} \quad \Omega_0 = \bm{k}\cdot\bm{v}_0.
\end{equation}

The former quantity is the parallel gradient operator and is an important quantity in the study of spectral stability \citepads{GoedbloedJP2019MoLa}. At the radius where $F$ vanishes, the Alfv\'en and slow continua both extend to the marginal frequency $\omega^2 = 0$. The magnetic tension is minimised at that location, counteracting less the driving forces of instability. \citetads{Suydam1958} derived a local criterion for such a point to be a clusterpoint where unstable discrete modes can accumulate at. These Suydam modes are highly localized to the radial surface. The generalisation of Suydam's criterion for a plasma with background flow was derived by \citetads{2004JPlPh..70..651W}.

The latter quantity in \cref{Eq: Fandds} is the frequency corresponding to the Doppler shift. From these equations it can be seen that every frequency $\omega$ becomes Doppler-shifted. This shift is along the real axis in the complex plane because the wavevector and flow are both real quantities. We use the following notation for the shifted frequency
\begin{equation}
    \tilde{\omega} = \omega - \Omega_0.
\end{equation}

Shifting the frequency is not the only alteration to the modes that the background flow might be responsible for. This can be seen from the right-hand-sides of the equations. In most equations, terms which are related to the background flow are present. 

We now reduce the set of Fourier-analysed MHD equations, \crefrange{Eq: MHDfourierfirst}{Eq: MHDfourierlast}, to two coupled first order differential equations; one for the total perturbed pressure and one for the radial velocity perturbation. This is a lengthy derivation, where in the first phase the equations are substituted into each other in favor for the perturbations $\rho$, $T$, $v_r$, and $Y$. In the second part the equations are reduced to two first-order differential equations, for $v_r$ and $Y$, by collecting, reshaping and eliminating terms. We finally obtain
\begin{align}\label{Eq: twodiffs1}
    &C_0 D(\tilde{\omega} Y) = - C_{1+} (\tilde{\omega} Y) + C_3 (rv_r),\\\label{Eq: twodiffs2}
    &C_0 D(rv_r) = C_2 (\tilde{\omega} Y) + C_{1-} (rv_r),
\end{align}

\noindent with the coefficients given by
\begin{align}\label{Eq: C0}
    C_0 =& ~r[\rho_0\tilde{\omega}^2 - F^2] C_t     ,\\
    \begin{split}
    C_{1\pm} =& ~C_1 \pm C_0 \frac{D(\Omega_0)}{\tilde{\omega}} \\
    =& -\frac{2m}{r} [FB_{0\theta} + v_{0\theta}\rho_0\tilde{\omega}] C_t \\
    &+ i \left[\tilde{\omega}\rho_0 + i(\gamma - 1)\left[\frac{\kappa_\parallel}{B_0^2}F^2 + \rho_0 \mathcal{L}_T\right]\right]  \\
    &\cross [(B_{0\theta}^2 - \rho_0v_{0\theta}^2)\tilde{\omega}^2 + [B_{0\theta}\tilde{\omega} + v_{0\theta}F]^2]\rho_0^2\tilde{\omega}^2 \pm C_0 \frac{D(\Omega_0)}{\tilde{\omega}},
    \end{split}\\
    \begin{split}
    C_2 =&  -ir^2\left[\frac{m^2}{r^2} + k^2\right]C_t \\
    &- [\tilde{\omega}\rho_0 + i(\gamma - 1)\left[\frac{\kappa_\parallel}{B_0^2}F^2 + \rho_0 \mathcal{L}_T\right]] r^2\rho_0^2 \tilde{\omega}^4,
    \end{split}\\
    \begin{split}    
    C_3 =& ~i\left[  [\rho_0\tilde{\omega}^2 - F^2]^2  + [\rho_0\tilde{\omega}^2 - F^2] rD\left[ \frac{B_{0\theta}^2}{r^2} - \frac{\rho_0v_{0\theta}^2}{r^2}  \right] \right.\\ 
    & \left. - \frac{4}{r^2}[FB_{0\theta} + \rho_{0}v_{0\theta}\tilde{\omega}]^2 \right]C_t  \\
    & -  \frac{1}{r^2} \left[\tilde{\omega}\rho_0 + i(\gamma - 1)\left[\frac{\kappa_\parallel}{B_0^2}F^2 + \rho_0 \mathcal{L}_T\right]\right] \\
    & \cross [(B_{0\theta}^2 - \rho_0v_{0\theta}^2)\rho_0\tilde{\omega}^2 + \rho_0[B_{0\theta}\tilde{\omega} + v_{0\theta}F]^2]^2,
    \end{split}\\
    \begin{split}\label{Eq: C_t}
    C_t =&  i\rho_0 \tilde{\omega}[\rho_0 \tilde{\omega}^2(\gamma p_0 +B_0^2)\\
    &  - \gamma p_0 F^2] + (\gamma-1)[\rho_0\tilde{\omega}^2 - F^2]\rho_0^3 \mathcal{L}_{\rho}\\
     & - (\gamma-1)\left[\frac{\kappa_{\parallel}F^2}{B_0^2} + \rho_0\mathcal{L}_{T}\right][\rho_0\tilde{\omega}^2(p_0+B_0^2) - p_0F^2].
    \end{split}
\end{align}

In the limiting case of vanishing background flow, the Doppler shift and flow related terms are neglected. \crefrange{Eq: twodiffs1}{Eq: twodiffs2} reduces to the set of equations derived by \citetads{1991SoPh..134..247V}, which in turn is formally identical to the equations obtained by \citetads{1974PhFl...17.1471A}, without any non-adiabatic terms. Note the difference in notation with \citetads{1991SoPh..134..247V}.


\subsection{Frieman-Rotenberg formulation}

One of the things to note from the previous equations, \crefrange{Eq: C0}{Eq: C_t}, is the $\pm$-sign in the coefficient $C_{1\pm}$. This coefficient consists out of the expression $C_1$ and a term due to the background flow. The second term arises only because of the use of a static Eulerian perturbation, $v_r$.

For spectral studies of equilibria with stationary flow a more appropriate formulation exists, the Frieman-Rotenberg formulation. A full description can be found in e.g. \citetads{1960RvMP...32..898F} and \citetads{GoedbloedJP2019MoLa}. A plasma displacement vector $\bm{\xi}$ is defined in a Lagrangian way, to connect the perturbed flow with the unperturbed flow. 



 

In this work we use the substitution from \citetads{2005A&A...444..337B}
\begin{equation}
    \bm{v}_1 = \pdv{\bm{\xi}}{t} + \bm{v}_0 \cdot \nabla \bm{\xi} - \bm{\xi} \cdot \nabla \bm{v}_0.
\end{equation}

For the radial component of the velocity perturbation, this yields
\begin{equation}
    v_r = -i\tilde{\omega}\xi_r,
\end{equation}

so there is only a contribution due to the Doppler shift. This is because the background flow is only dependent on the radial coordinate.

Using this substitution, the set of two first-order differential equations are given by   
\begin{align}\label{Eq: 1orderdiffs_FR}
    &C_0 D(Y) = - C_{1} Y - iC_{3} (r\xi_r),\\ \label{Eq: 1orderdiffs_FR2}
    &C_0 D(r\xi_r) = iC_{2} Y + C_{1} (r\xi_r),
\end{align}

The benefit of this notation is that the cumbersome $C_{1\pm}$ just becomes $C_1$, dropping the $\pm C_0\frac{D(\Omega_0)}{\tilde{\omega}}$.

These two first-order differential equations can be combined into one second-order differential equation. Substituting \cref{Eq: 1orderdiffs_FR} into \cref{Eq: 1orderdiffs_FR2} yields after some manipulation
\begin{equation}\label{Eq: 2orderdiff}
    D\left[ \frac{C_0}{C_2} D(r\xi_r) \right] - \left[ \frac{C_1^2 + C_2C_3}{C_0C_2} + D\left[ \frac{C_1}{C_2}\right]  \right] r\xi_r = 0.
\end{equation}

This second-order differential equation is formally the same as obtained by \citetads{1958ZNatA..13..936H}. The coefficients are different because they considered a static, adiabatic plasma. This equation is used in \cref{Sec: discretemodes}. Neglecting the non-adiabatic effects, \crefrange{Eq: 1orderdiffs_FR}{Eq: 1orderdiffs_FR2}, are formally identical and reduce to the set of equations derived by \citetads{1987PhFl...30.2167B}, \citetads{1992SoPh..138..233G} and \citetads{2004JPlPh..70..651W}.

\subsection{The continua}

The set of first-order differential equations, \crefrange{Eq: twodiffs1}{Eq: twodiffs2}, becomes singular as the coefficient $C_0$ vanishes, as discussed by \citetads{1974PhFl...17.1471A}. The eigenfunctions of such continuum solutions are non-square integrable. These are of special interest because of their local character, leading in the case of thermal instabilities to rapid, in situ condensation formation. The coefficient $C_0$ is given by
\begin{equation}
    C_0 = r[\rho_0\tilde{\omega}^2 - F^2] C_t = 0.
\end{equation}

The first possibility for this coefficient to vanish is when the term in the brackets vanishes. This leads to the forward and backward propagating Alfv\'en continua. The Alfv\'en modes are not influenced by the non-adiabatic effects. The continuum modes are real and resonate locally, meaning that the Alfv\'en speed varies with $r$ and at every point you can define a continuum mode as  
\begin{equation}\label{Eq: alfvencont}
    \tilde{\omega}_{A\pm}(r) = \pm\frac{F(r)} {\sqrt{\rho_0(r)}}.
\end{equation}

The Alfv\'en continuum is thus a set of frequencies where each frequency depends on a certain radius. Those modes are altered by the background flow. However, this is just a shift along the real axis and does not alter the stability of the modes. 

The second way how the coefficient $C_0$ can vanish, is when the coefficient $C_t$ vanishes. This is the thermal coefficient and shown here again for clarity:
\begin{equation}\label{Eq: Ctcont}
    \begin{split}
        C_t(\tilde{\omega},r)  =  &~i\rho_0 \tilde{\omega}[\rho_0 \tilde{\omega}^2(\gamma p_0 +B_0^2) - \gamma p_0 F^2]\\
        & - (\gamma-1)\left[\frac{\kappa_{\parallel}F^2}{B_0^2} + \rho_0\mathcal{L}_{T}\right][\rho_0\tilde{\omega}^2(p_0+B_0^2) - p_0F^2]\\
        & + (\gamma-1)[\rho_0\tilde{\omega}^2 - F^2]\rho_0^3 \mathcal{L}_{\rho} = 0.
        \end{split}
\end{equation}
This equation is a third-order polynomial in frequency and can be solved for every radial coordinate. The three solutions are the thermal continuum and the two, forward and backward propagating, slow continua. The slow continua are modified by the non-adiabatic effects and can be unstable or damped based on the physical properties of the background equilibrium. The thermal continuum was first described by \citetads{1991SoPh..134..247V}. It is the extension of the thermal mode discussed by \citetads{1965ApJ...142..531F} for an non-uniform medium. The thermal continuum is quite useful, since it allows one to easily determine the stability of a plasma, disregarding discrete modes for the moment, without solving the set of Fourier-analysed MHD equations numerically. 

When the non-adiabatic effects are neglected \cref{Eq: Ctcont} reduces to a second-order polynomial, which holds only the slow continua. When also flow is neglected the equation than reduces to the results in for example the review paper by \citetads{1984PhyD...12..107G}.

Now what is the effect of the background flow on the coefficient $C_t$ and by extension on the slow and thermal continua? The equation has the same functional shape as the equation obtained by \citetads{1991SoPh..134..247V}, taking into account the difference in notation. However, it is not to be solved for just the frequency. Here it needs to be solved for the frequency modified by the Doppler shift. Hence the solutions of the third-order polynomial are shifted in the complex plane. Since the Doppler shift is a shift along the real axis, it does not alter the stability. So modes that are unstable in a given equilibrium are also unstable in the analogous, flow-included equilibrium, as long as the flow does not change the equilibrium itself.

\section{Numerical investigation of the continua}\label{Sec: num_inv_cont}

In the previous section we derived the equations for the continua and investigated the influence of a stationary background flow on the thermal continuum in particular. In this section we confirm our analytic results numerically using the eigenvalue code \textit{Legolas} \citepads{2020ApJS..251...25C,2022JPlPh..88c9021D,2023arXiv230710145C}. 

We start with the description of the numerical setup which we use to represent a solar tornado in the solar corona. Secondly, we investigate the MHD spectrum of such an equilibrium. Furthermore, we vary several parameters of the equilibrium, which include the azimuthal velocity profile of the background flow and the optically thin radiative cooling curve. Lastly, we study thermal conduction and how different wavenumbers influence the stability of the thermal continuum.

\subsection{Numerical setup}\label{Sec: Setup_TIflow}

The numerical code used in this work is \textit{Legolas} \citepads{2020ApJS..251...25C,2022JPlPh..88c9021D,2023arXiv230710145C}. It is an open-source code designed to solve the set of linearised MHD equations to perform spectral analysis of the plasma. While relying on a finite element representation, the Fourier-analysed equations are transformed into a generalised eigenvalue problem of the form
\begin{equation}
     A\bm{x} = \omega B\bm{x}
\end{equation}

\noindent where $A$ and $B$ are matrices containing equilibrium variations. The vector $\bm{x}$ denotes the state vector of perturbed quantities. \textit{Legolas} is written in Fortran and modularised. It therefore has a very large range of physical effects implemented. For the work considered here, only background flow, optically thin radiative cooling, and thermal conduction are needed. However, other effects, such as resistivity and external gravity, are also available. Its post-processing Python package, \textit{pylbo}, allows for easy access to the eigenvalues and eigenfunctions, and to the details of the equilibrium. \textit{Legolas} supports 3D cylindrical and cartesian geometries with one dimensional variation.

The equilibrium that we use is a cylindrical plasma with one dimensional variation that represents a solar tornado. Magnetic models of solar tornadoes exist in the literature \citepads{2015ApJ...808L..23L,2018ApJ...863..147L,2018PhPl...25e4503O}. Since our main result is independent of the profiles of the equilibrium parameters, we base our model on the more simple and well-known Gold-Hoyle equilibrium, first described by \citetads{1960MNRAS.120...89G}. This model was also used by \citetads{1991SoPh..134..247V}, giving us the opportunity for direct comparison. 

The magnetic field of a Gold-Hoyle equilibrium is given by
\begin{align}
    B_{0\theta} &= \frac{B_{0c} \mu r}{1+ \mu^2 r^2},\\
    B_{0z} &= \frac{B_{0c}}{1+\mu^2 r^2},
\end{align}

\noindent where $B_{0c}$ is the total magnetic field strength at the axis. In these equations $\mu$ is the inverse pitch and is related to the magnetic field components via
\begin{equation}
    \mu = \frac{B_{0\theta}}{rB_{0z}}.
\end{equation}

\noindent The magnetic field is helical, just as we want for a magnetic tornado model. The parameter $\mu$ determines how twisted the field is. We consider a constant pitch. One of the benefits of using the Gold-Hoyle magnetic field is that it is force-free.

The background flow is also helical with the profiles taken as
\begin{align}
    v_{0\theta} &= v_{0\theta b} r^\alpha,\\
    v_{0z} &= v_{0z},
\end{align}

\noindent where $v_{0\theta b}$ is the azimuthal velocity at the boundary at $r=1$ and $\alpha$ is the exponent determining the shape of the azimuthal velocity profile. The axial velocity is taken to be constant throughout the cylinder. 

The pressure, temperature, and density are coupled due to the ideal gas law. If we take the density to be constant, the temperature cannot be constant, with a non-constant pressure. We assume a radially dependent density profile in the equilibrium when using \textit{Legolas}
\begin{equation}
    \rho_0 = \rho_{0b}  r^\tau,\\
\end{equation}

\noindent with $\rho_{0b}$ the density at the boundary and $\tau$ being the exponent. However, we keep $\tau$ fixed to zero to effectively work with a constant density. Unlike the real Gold-Hoyle equilibrium the pressure is not constant. The pressure is fixed by the mechanical equilibrium of our flow-included setup, \cref{Eq: Mechequi}. As the magnetic field is force-free, it becomes a relation between the thermal pressure and the azimuthal velocity. The thermal pressure needs to balance the centrifugal force. Using the ideal gas law and the profiles of aforementioned equilibrium quantities the profile of the background temperature is given by
\begin{equation}
    T_0 = \frac{v_{0\theta b}^2}{2\alpha + \tau} r^{2\alpha} + T_{0c} r^{-\tau},
\end{equation}

\noindent where $T_{0c}$ is the background temperature at the axis when the density is constant. In the limiting case of vanishing background flow and constant density, this simplifies to the literature Gold-Hoyle equilibrium of \citetads{1960MNRAS.120...89G}.

We consider a domain for the dimensionless radial coordinate $r$ from 0 to 1. The normalisation used in \textit{Legolas} is described in \citetads{2020ApJS..251...25C}. We chose to set the unit temperature, unit magnetic field, and unit length scale to $10^6$\,K, $10$\,G, and $10^9$\,cm, respectively. We consider an electron-proton plasma with a mean molecular weight of 0.5. 

The default values for the physical parameters are given in \cref{Tab: setupTIflow}. We consider a coronal plasma with a background magnetic field strength of 10 G. The plasma beta of the medium is approximately 0.1, which is typical for prominences in the solar corona \citepads{2018LRSP...15....7G}. For the background flow we only consider velocities in the order of a few tens of kilometers, as observed for solar tornados by use of solar spectroscopy \citepads{2018ApJ...852...79Y,2021A&A...653A..94B}. We initially use a $r$-squared profile for the azimuthal velocity. With our choice of values for the parameters, we consider a weak inhomogeneity. This ensures that the continua are well separated in the complex plane. The equilibrium values are chosen such that discrete modes corresponding to well-known instabilities, like Kelvin-Helmholtz ones which are not the topic of this investigation, are not present.

\begin{table}[hbt!]
    \caption{The values for the physical quantities used in the benchmark case. $B_{0c}$ is the magnetic field strength at the axis. $\mu$ is the inverse pitch of the magnetic field. $\rho_{0b}$, $\tau$, and $T_{0c}$ denote the density at the boundary, the exponent of the density profile, and the temperature at the axis, respectively. Lastly the flow-related quantities are $v_{0\theta b}$, $\alpha$, and $v_{0z}$ and represent the azimuthal velocity at the boundary, the exponent of the azimuthal profile, and the constant axial velocity, respectively.}              
    \label{Tab: setupTIflow}     
    \centering                                 
\begin{tabular}{c c}          
    \hline\hline      
    Quantity & Value \\ 
    \hline                                   
     $B_{0c}$ & 10 G  \\
     $\mu$ &  0.1 \\
     $\rho_{0b}$ & 2.3 $\cdot$ 10$^{-15}$ g cm$^{-3}$ \\
     $\tau$ & 0 \\
     $T_{0c}$ & $10^6$ K \\
     $v_{0\theta b}$ &  10$^5$ cm s$^{-1}$ \\
     $\alpha$ & 2 \\
     $v_{0z}$ &  10$^5$ cm s$^{-1}$ \\
     \hline
\end{tabular}
\end{table}

\begin{figure}[htbp]
    \centering
    \resizebox{\hsize}{!}{\includegraphics{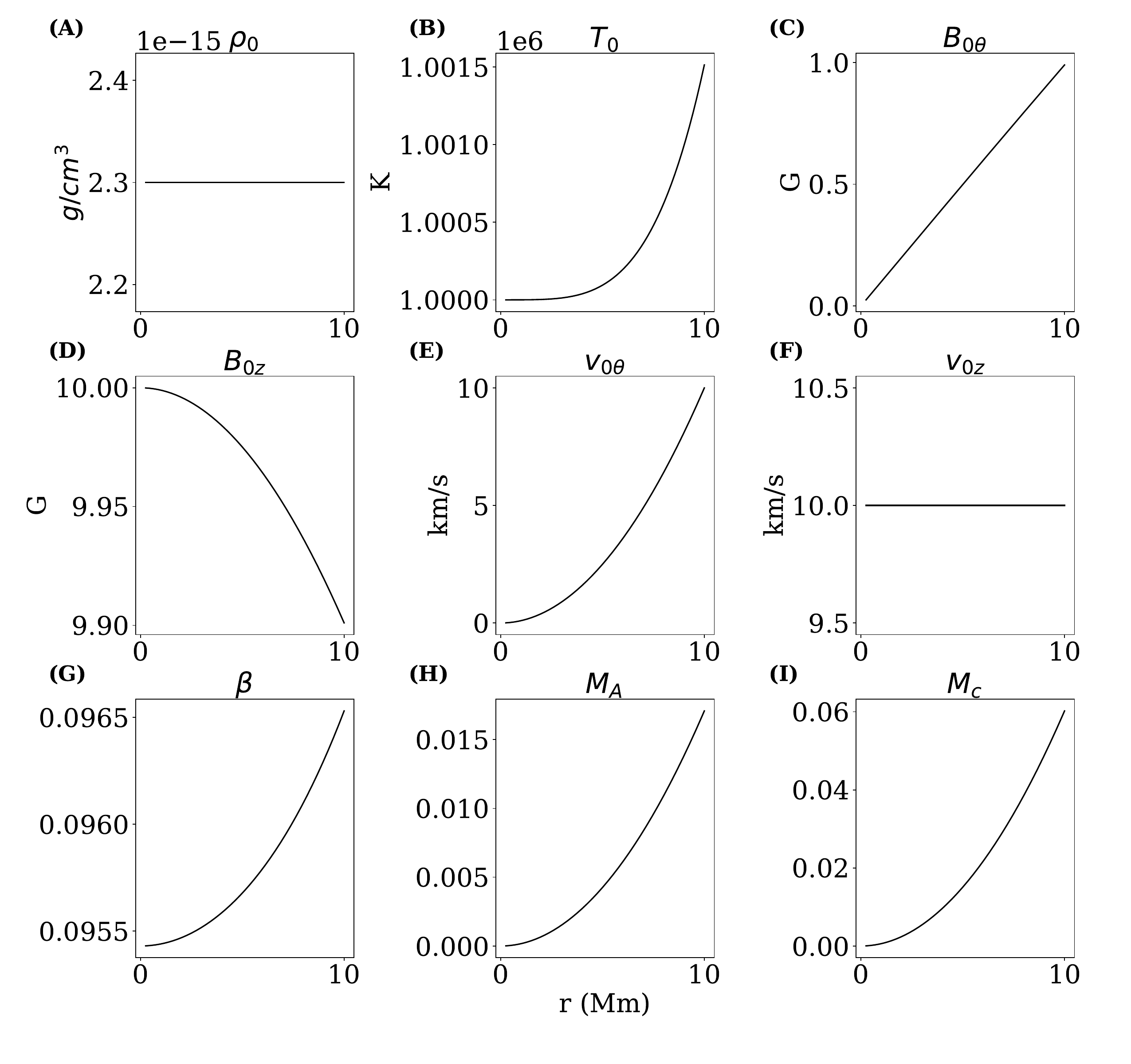}}
    \caption{The equilibrium profiles of the benchmark setup with respect to the radial coordinate. From left to right, top to bottom the quantities are: the density, the temperature, the azimuthal magnetic field, the axial magnetic field, the azimuthal velocity, axial velocity, the plasma beta, the Alfv\'en mach number, and the standard mach number with respect to the sound speed.}
    \label{Fig: Equilibriasetup}
\end{figure}

\begin{figure}[htbp]
    \centering
\includegraphics[width=.485\linewidth]{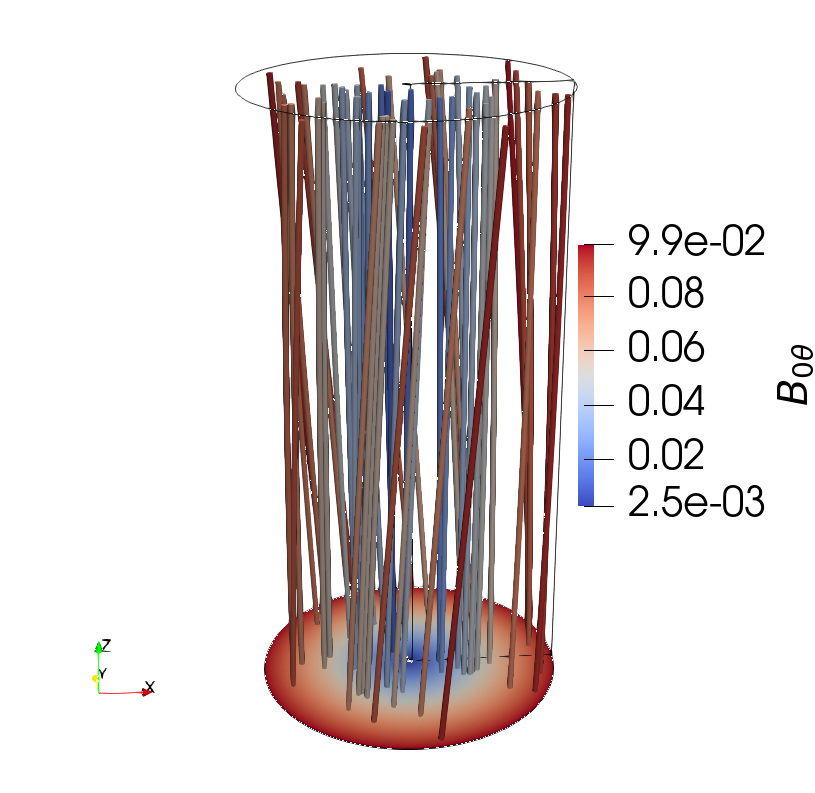}
    \hfill
\includegraphics[width=.485\linewidth]{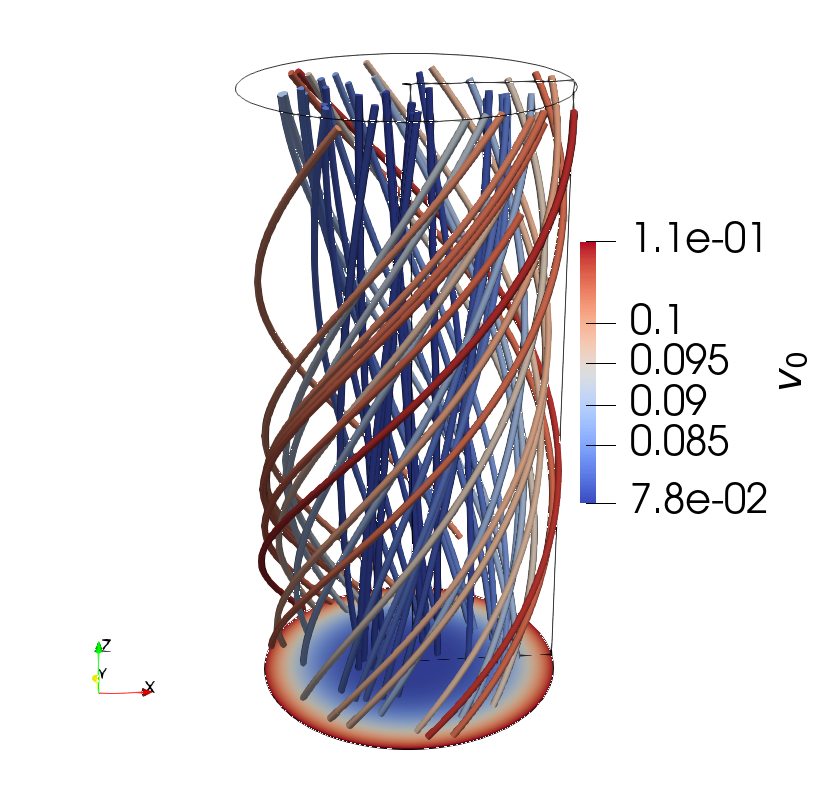}
    \caption{The magnetic and velocity field lines of the background equilibrium are shown left and right, respectively. The magnetic field lines are coloured according to the magnitude of the $B_{0\theta}$ component. The velocity field lines take the colour of their magnitude.} 
    \label{Fig: 3Dbench}
\end{figure}

In \cref{Fig: Equilibriasetup} several of the background parameters and some profiles of important derived quantities are shown. In panel (A) the constant density is shown. The temperature profile is shown in panel (B). The temperature is not constant, because the pressure balances the centrifugal force. It slightly increases outwards, but remains close to $10^6$ Kelvin. The panels (C) and (D) show the azimuthal and axial magnetic field components, respectively. The magnetic field is mostly vertical, as to be expected for vertical columns. The azimuthal field is an order of magnitude weaker than the axial field. The medium is dominated by the magnetic forces as can be seen from the plasma beta shown in panel (G). The plasma beta is around 0.096 and spatially varies a little bit because of the variations in the pressure and magnetic field. The velocity profiles are shown in panels (E) and (F) of \cref{Fig: Equilibriasetup}. The azimuthal velocity is quadratically increasing, while the axial velocity is constant. From panels (H) and (I), it can be seen that the azimuthal velocity is always subsonic and sub-Alfv\'enic, via the corresponding Mach numbers. The sound speed and Alfv\'en speed are around 165 km\,s$^{-1}$ and 587 km\,s$^{-1}$ for the given parameters, respectively. \cref{Fig: 3Dbench} depicts the magnetic and velocity field lines of the stationary background in 3D. The magnetic field lines are indeed predominantly vertical, only the outermost have more twist. They are coloured according to the magnitude of the $B_{0\theta}$ component, highlighting the helicity. The velocity field lines are more twisted, whereas the lines at the center are mostly straight. The velocity increases going outwards, due to the increase in azimuthal flow.

For all cases, the dimensionless wavenumbers, $m$ and $k$, are both set to unity, unless stated otherwise. One cannot investigate thermal instability without radiative cooling. \textit{Legolas} handles this using optically thin radiative cooling tables. Commonly used cooling tables are described by \citetads{2021A&A...655A..36H} and in their appendix. In the default case we use the SPEX\_DM cooling curve \citepads{2009A&A...508..751S}. The cooling curve is one of the parameters that we vary in the following sections. Thermal equilibrium needs to hold at the initial state. A constant background heating, equal to the radiative heating at $t=0$, is set in order to facilitate this. Thermal conduction is not used in the benchmark case. However, we use it later on, but only parallel conduction.

We use \textit{Legolas 2.0}, the most recent version. This second version was recently discussed in detail by \citetads{2023arXiv230710145C}. \textit{Legolas} has several solvers at its disposal. They have become more accurate and faster in the second version. Solvers, such as the default \textit{QR-invert}, calculate the whole spectrum of eigenmodes. Other solvers, such as the \textit{Arnoldi} solvers, are tailored to finding a certain amount of specific modes. We typically use the default \textit{QR-invert} solver, but have used the \textit{Arnoldi} solvers when in doubt of the convergence of the modes. For details about the different solvers in \textit{Legolas} we refer to \citet{2023arXiv230710145C}. We used 500 gridpoints in the base grid to keep the memory usage and computation time in check. The only available boundary condition is a perfectly conduction wall at $r = 1$.

\subsection{Benchmark}

\subsubsection{The adiabatic case}\label{Sec: adiabatic_bench}

\begin{figure}[htbp]
    \centering
    \resizebox{\hsize}{!}{\includegraphics{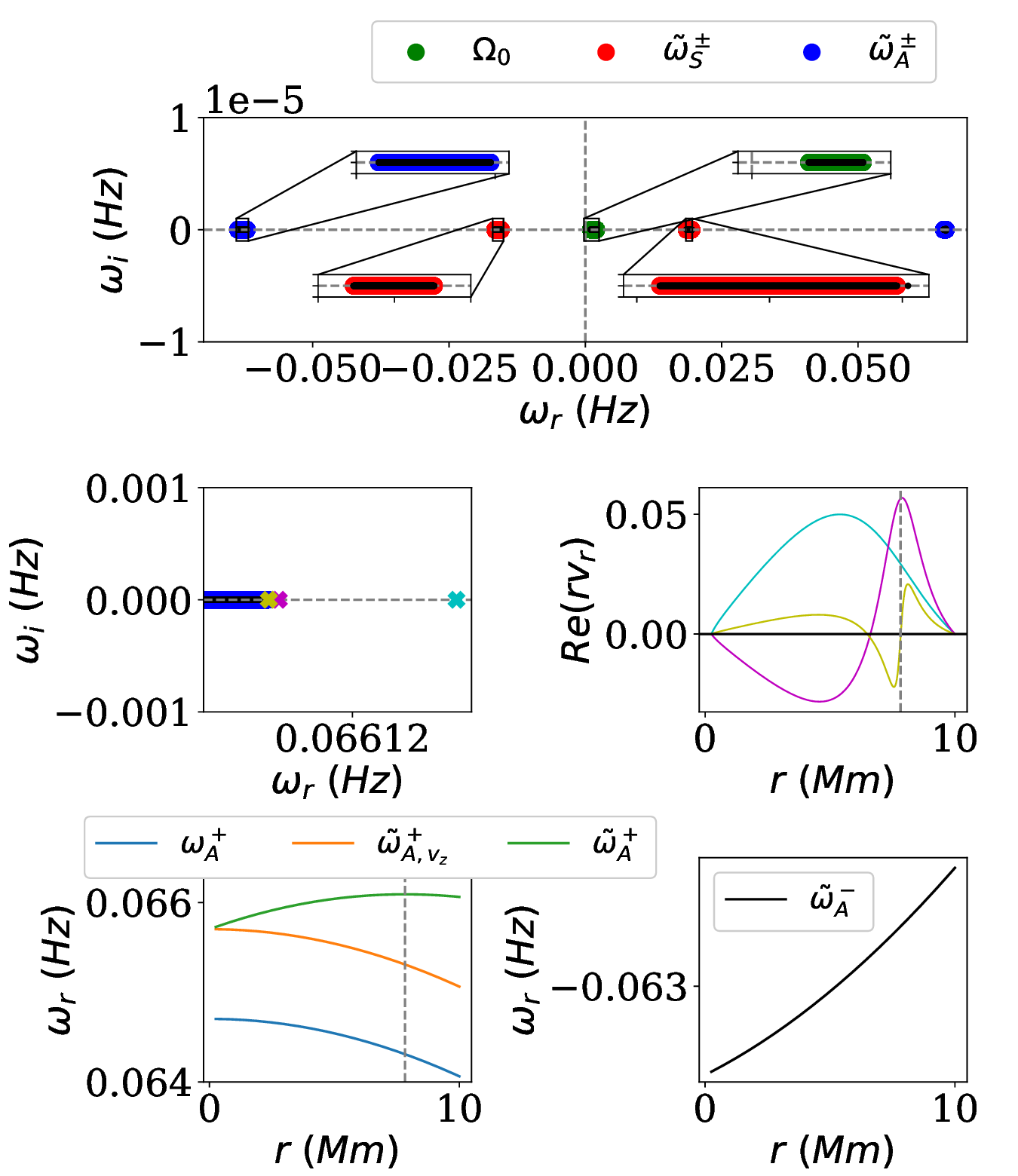
    }}
    \caption{In the top panel the overview spectum of the adiabatic benchmark equilibrium with both azimuthal and axial flow is given. Insets zooming in on the continua are added. The middle left panel is a zoom-in onto the upper edge of the forward-propagating Alfv\'en continuum. Three discrete Alfv\'en modes are marked. The corresponding eigenfunctions are shown in the middle right panel. The bottom left panel shows the forward-propagating Alfv\'en continuum as a function of radius for the adiabatic benchmark equilibrium, for the static version of the benchmark equilibrium, and for a version of the benchmark equilibrium with only axial flow. The backward-propagating Alfv\'en continuum is shown in the bottom right panel.}
    \label{Fig: flow_adia_overview}
\end{figure}

We first take a look at the adiabatic spectrum in order to understand some peculiarities that arise due to the background flow, regardless of non-adiabatic effects. The spectrum of the benchmark equilibrium, a Gold-Hoyle magnetic field with a constant axial and $r$-squared azimuthal flow profile, is given in the top panel of \cref{Fig: flow_adia_overview}. There are insets showing the two slow continua, the backward-propagating Alfv\'en continuum, and the Doppler continuum. The continua are spread out along the real axis according to their intrinsic resonances and the Doppler shift. There is a single discrete slow mode next to the upper edge of the forward-propagating slow continuum. No discrete modes appear next to the backward-propagating slow continuum or the backward-propagating Alfv\'en continuum.

It is interesting to note that the azimuthal background flow facilitates the clustering of a series of discrete Alfv\'en modes to the upper edge of the forward-propagating Alfv\'en continuum. The discrete modes do not appear in the static case or when only including axial flow in the background equilibrium. The clustering can be seen in the middle left panel of \cref{Fig: flow_adia_overview}. The three discrete modes are marked with coloured crosses and their eigenfunctions are shown in the middle right panel in the same colours. It is an anti-Sturmian sequence as the frequency reduces with the number of nodes. Discrete modes are known to cluster towards extrema of continua \citepads{1974PhFl...17..908G}. To explain the appearance of these discrete Alfv\'en modes the forward-propagating Alfv\'en continuum is plotted in the bottom left panel together with the Alfv\'en continuum of a static version of the equilibrium and that of an equilibrium that only includes axial flow. The bottom right panel shows the backward-propagating Alfv\'en continuum of the general flow case discussed here. From the former panel it can be seen that only in the case of both azimuthal and axial flow there is an internal maximum. In the case with only axial flow the continuum is still monotonously decreasing, but it is increased with a constant Doppler shift since the axial velocity is taken constant. The inclusion of the azimuthal flow, not linearly dependent on the radius, causes the Doppler shift to be non-constant, leading to this maximum. The backward-propagating Alfv\'en continuum also has no internal extremum because the addition of the Doppler shift to the continuum is not in the correct radial position to create it. \citetads{2004JPlPh..70..651W} derived a criterion for clustering towards extrema of slow and Alfv\'en continua using a Frobenius expansion. When a factor dependent on mode numbers and equilibrium and variations, $D_A^+$, evaluated at the location of the extremum is larger than 0.25, discrete modes arise. For the Alfv\'en continuum considered here we have $D_A^+(r= 0.78) = 1.58$. This is larger than 0.25, and hence we would indeed expect discrete Alfv\'en modes to cluster towards the maximum of the forward-propagating Alfv\'en continuum.

\subsubsection{The non-adiabatic case}

We now extend on the literature results by investigating the spectrum of a plasma both influenced by non-adiabatic effects and background flow. We still neglect thermal conduction for the time being.

\begin{figure}[htbp]
    \centering
    \resizebox{\hsize}{!}{\includegraphics{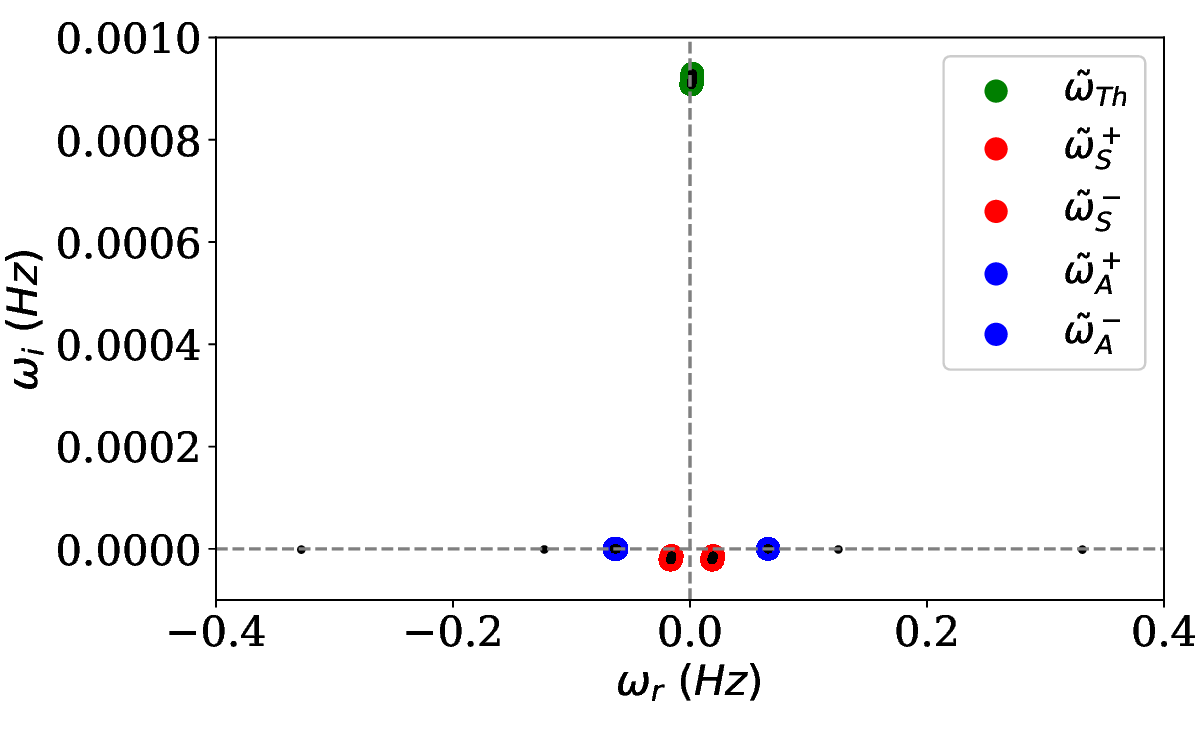}}
    \caption{The spectrum of the benchmark equilibrium. The black dots are the results obtained using \textit{Legolas}. The green, red, and blue dots are the thermal, slow and Alfv\'en continua, respectively.}
    \label{Fig: flow_nonadia_overview}
  \end{figure}

\begin{figure}[htbp]
    \centering
    \resizebox{\hsize}{!}{\includegraphics{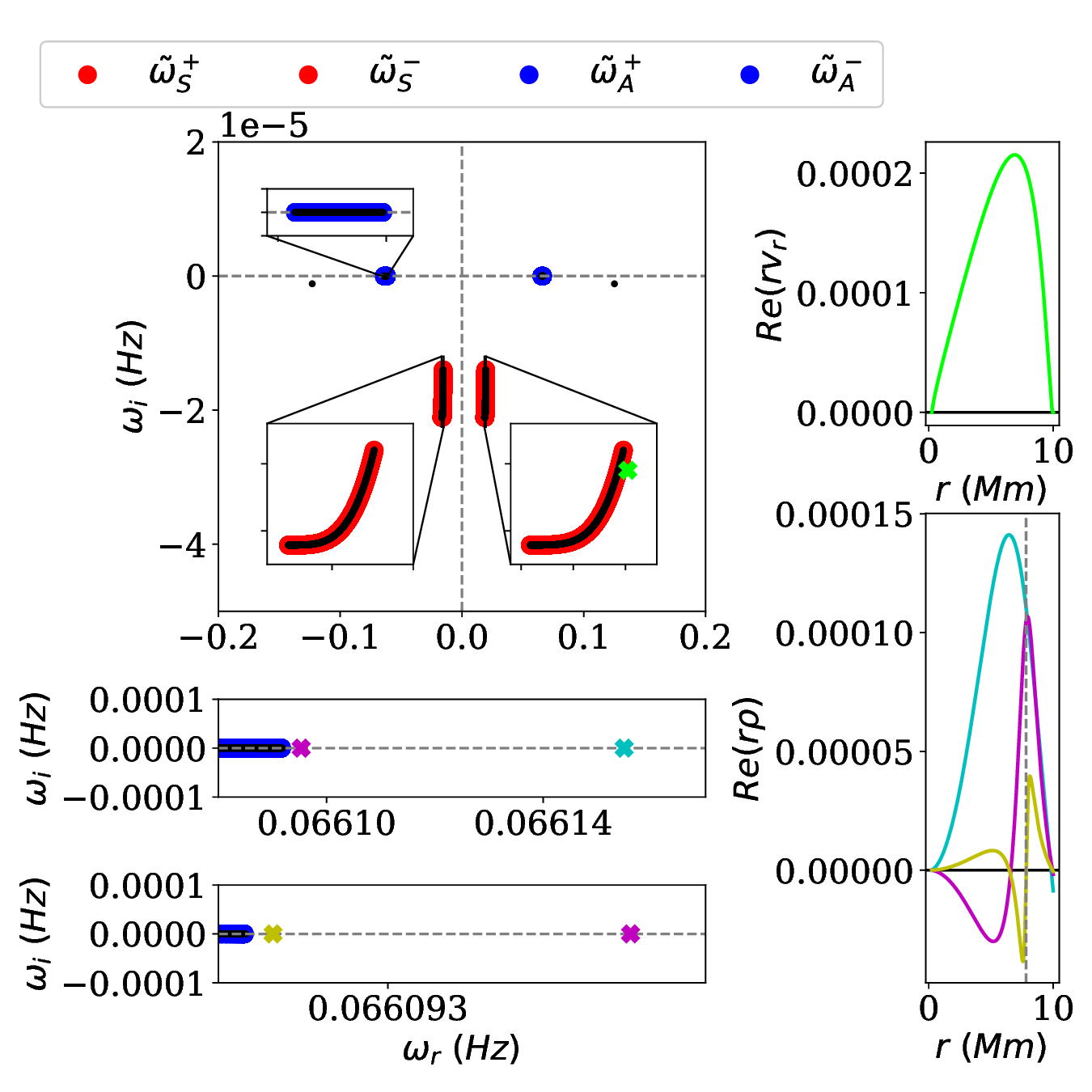}}
    \caption{The slow and Alfv\'en continua of the benchmark equilibrium are depicted in red and blue, respectively. The results obtained with \textit{Legolas} are the black dots. In the top-left panel, there are insets zooming into the backward propagating Alfv\'en continuum and the two slow continua. The two panels below the main spectrum are zoom-ins onto the forward progating Alfv\'en continuum, with another zoom towards the right edge. The discrete modes are marked with coloured crosses. The eigenfunction of the discrete slow mode is shown in the top-right panel. In the bottom-right panel the eigenfunctions of the three discrete Alfv\'en modes are shown together with a grey dotted line denoting the location where the Alfv\'en continuum reaches its maximum.}
    \label{Fig: flow_nonadia_slowalfven}
\end{figure}

The spectrum showing the overview of the modes is given in \cref{Fig: flow_nonadia_overview}. The black dots are the results of the \textit{Legolas} run. The blue, green, and red dots are the analytic expressions for the Alfv\'en, thermal, and slow continuum modes, respectively, and correspond to equations \cref{Eq: alfvencont,Eq: Ctcont}. For the used equilibrium the thermal continuum is unstable, while the slow continua are damped. The Alfv\'en continua remain real and stable, as to be expected. The beginning of the fast branches are visible at the outsides of the Alfv\'en continua. Two modes are shown on each side. Those branches extend to infinity and are damped, as can be seen more clearly in the zoomed-in \cref{Fig: flow_nonadia_slowalfven}.

It is also clear from \cref{Fig: flow_nonadia_overview} that the results obtained with \textit{Legolas} match perfectly with the analytic expressions for the continua.

\begin{figure}[htbp]
    \centering
    \resizebox{\hsize}{!}{\includegraphics{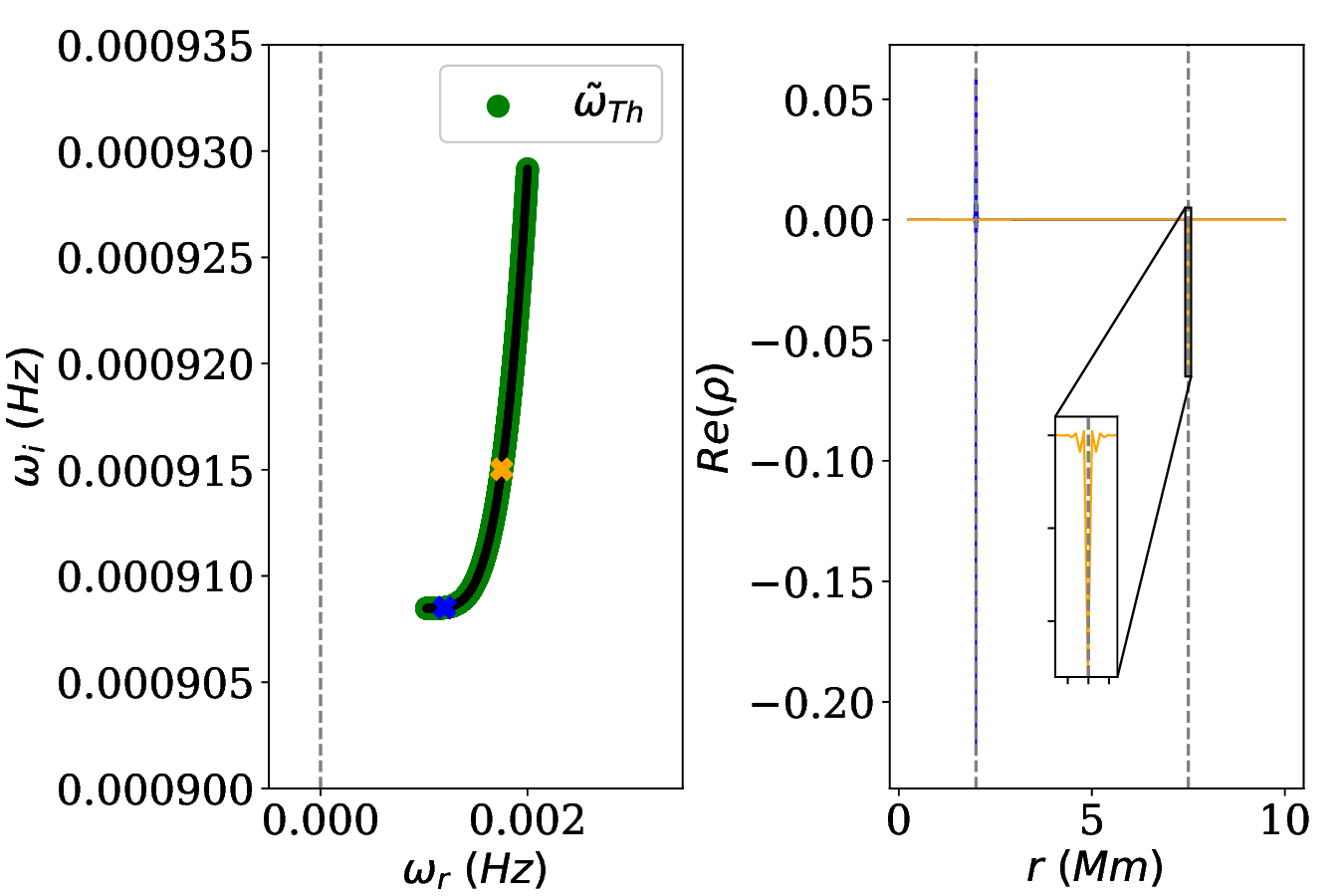}}
    \caption{The left panel shows a zoom-in of the thermal continuum of the benchmark equilibrium with the vertical axis as a grey dotted line. The density eigenfunctions of the orange and blue modes marked by crosses in the left panel are shown in the right panel. An inset focused on the localised non-square integrable shape is given in the right panel. In the right panel, the vertical grey lines indicate the radial position of the resonance.}
    \label{Fig: flow_nonadia_thermal}
\end{figure}

In the left panel of \cref{Fig: flow_nonadia_thermal} the thermal continuum demonstrates that the numerical result matches perfectly with the analytic green curve. A background flow Doppler shifts the thermal continuum as can be seen from the analytical expression, \cref{Eq: Ctcont}, derived earlier in this work. This shift is along the real axis. In the right panel of \cref{Fig: flow_nonadia_thermal} the density eigenfunctions of two continuum modes, denoted with yellow and blue crosses in the left panel, are depicted. The eigenfunctions are very localised and show a non-square integrable shape, as can be seen most easily in the inset.

The thermal continuum has a curved shape in the complex plane. At first glance, one might think that the shape of the shifted continuum is due to the chosen velocity profiles. However this is not the complete truth. Since we are working in a cylindrical coordinate system, the Doppler shift is not just the squared shape of the azimuthal velocity profile. It is also influenced by the axial velocity profile and the wavenumbers, such that
\begin{equation}\label{Eq: Dopplershift}
    \Omega_0 = \bm{k} \cdot\bm{v}_0 = \frac{m}{r}v_{\theta 0} + k v_{z0}.
\end{equation}

\begin{figure}[htbp]
    \centering
    \resizebox{\hsize}{!}{\includegraphics{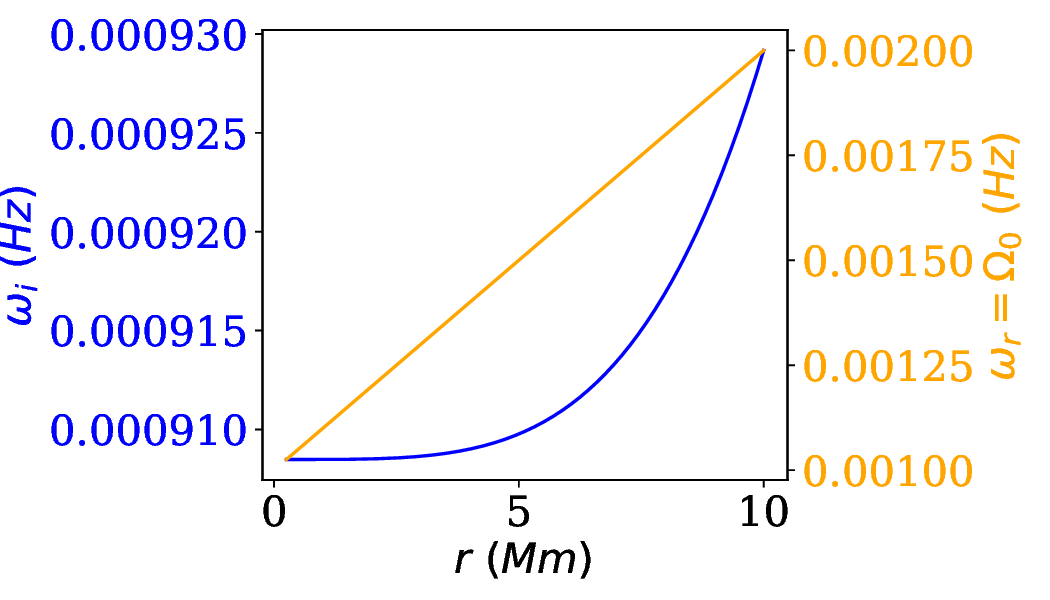}}
    \caption{The imaginary and real parts of the thermal continuum of the benchmark equilibrium as a function of radius. They are shown as blue and orange curves, respectively.}
    \label{Fig: flow_nonadia_thermal_split}
\end{figure}

In the case of a squared profile for the azimuthal velocity and a constant axial velocity, the Doppler shift is linear. This can be seen on \cref{Fig: flow_nonadia_thermal_split}, where the real and imaginary parts of the thermal continuum modes are plotted with respect to the radial coordinate. If the shape was dominated by the Doppler shift, one would expect just a straight line in the complex plane. The curved shape of the continuum is composed out of the real and imaginary part. Hence, the distribution of the imaginary part of the continuum in space is of relevance. This also has a curved shape as can be seen by the blue curve in \cref{Fig: flow_nonadia_thermal_split}. About half of the continuum modes, the ones corresponding to radii less than 5 Mm, have relatively low and similar growth rates. Due to the linear shift of those modes with nearly the same growth rate, an approximately horizontal part of the thermal continuum is obtained. The continuum modes, which are localised most outwards, are the most unstable and experience the largest Doppler shift. The fact that the outermost modes are more unstable than the innermost modes, can also be seen in \cref{Fig: flow_nonadia_thermal}. The equilibrium parameters, set in \cref{Sec: Setup_TIflow} and shown in \cref{Fig: Equilibriasetup}, determine how the imaginary part of the thermal continuum varies with radius. This can change drastically with only a minor change to any of the equilibrium profiles, which is discussed further.

The forward and backward propagating slow and Alfv\'en continua are shown in more detail in the top-left panel of \cref{Fig: flow_nonadia_slowalfven}, denoted by their typical colours of red and blue, respectively. For both types of continua the numerical results match the analytic expressions. The slow continua are governed by \cref{Eq: Ctcont}, which is the same equation as the thermal continuum. The Alfv\'en continuum is spread out according to \cref{Eq: alfvencont}. Besides the continua there are some more things to note. Just as in the adiabatic case shown in \cref{Sec: adiabatic_bench}, the inclusion of a background flow creates a maximum in the forward propagating Alfv\'en continuum. Three discrete modes can be seen to cluster towards the right edge in the bottom panels of \cref{Fig: flow_nonadia_slowalfven}. The modes are marked with different colours and are very slightly damped, contrary to their completely real nature in the adiabatic case. This is due to the non-adiabatic effects. The damping of discrete Alfv\'en modes by non-adiabaticity has been shown by \citetads{1993SoPh..144..267K}. The eigenfunctions are shown in the bottom-right panel. The grey dotted line denotes the location of the maximum in the Alfv\'en continuum. It can be seen that the number of nodes increases for modes closer to the right edge of the continuum, i.e. with decreasing frequency. This is thus an anti-Sturmian sequence, as to be expected. The backward-propagating Alfv\'en continuum has no extremum and no discrete modes cluster towards its edges. There also appears a discrete slow mode next to the forward propagating slow continuum. It is marked by a green cross and its eigenfunction is shown in the top-right panel. The fast modes, of which one can be seen on either side in the overview panel in the top-left of \cref{Fig: flow_nonadia_slowalfven}, are also damped.

\subsection{Azimuthal velocity profile}

One of the important and interesting parameters of a background flow is the azimuthal velocity profile. Here we compare two different profiles with the benchmark case discussed previously. We keep the axial velocity profile constant and do not change the magnitudes.

\begin{figure}[htbp]
   \centering
   \resizebox{\hsize}{!}{\includegraphics{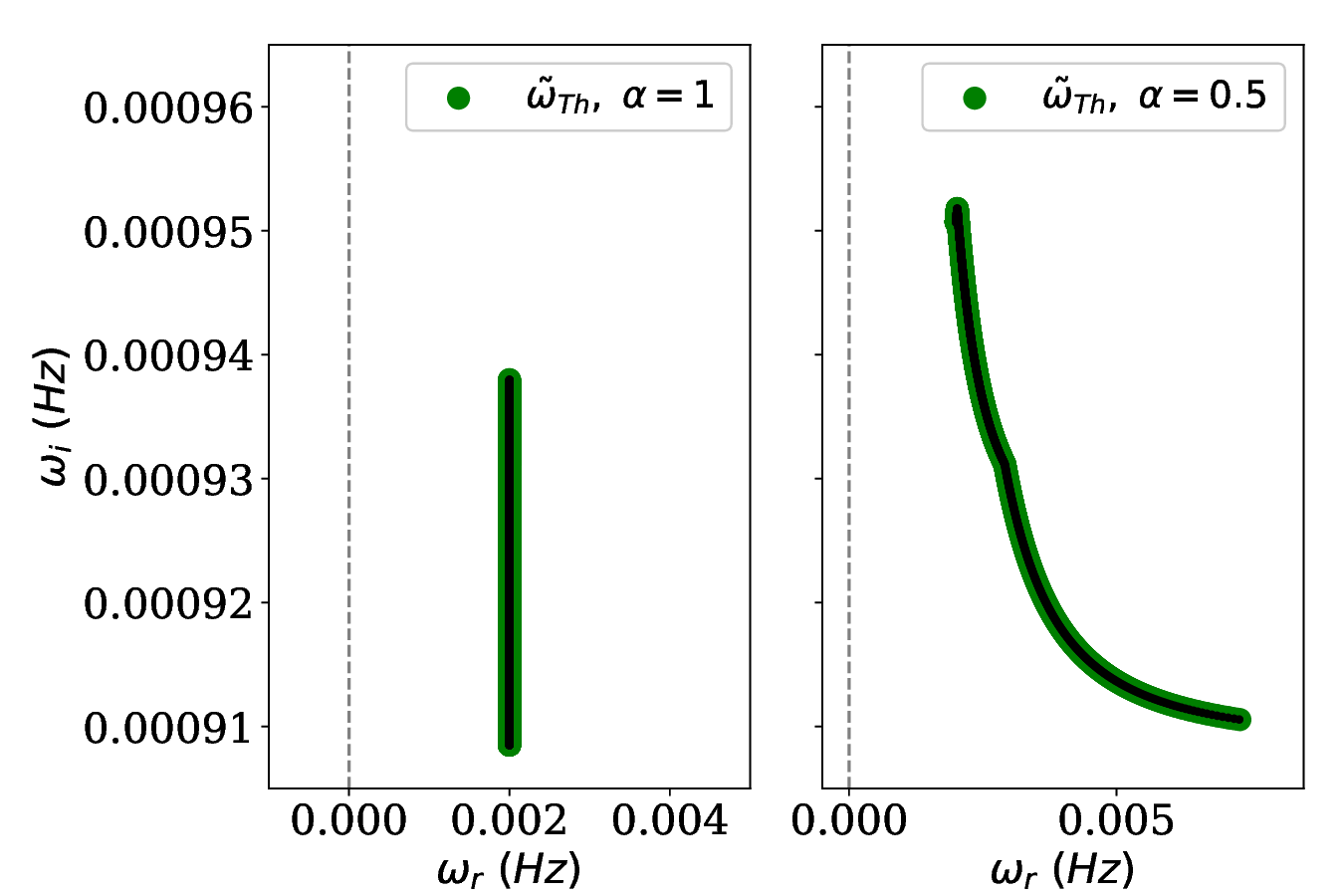}}
   \caption{The left and right panels shows the thermal continuum for the $\alpha = 1$ and $\alpha = 0.5$ cases, respectively. The analytic expressions are shown in green, while the \textit{Legolas} results are the black dots.}
   \label{Fig: flow_nonadia_thermal_alpha}
\end{figure}

First, we consider the basic case of an azimuthal velocity given by 
\begin{equation}
    v_{0\theta} = v_{0\theta b} r,
\end{equation}

\noindent which has a simple linear dependency on the radius, with $\alpha = 1$. This linear dependence on radius means that the Doppler shift, calculated with \cref{Eq: Dopplershift}, is constant. In the left panel of \cref{Fig: flow_nonadia_thermal_alpha}, the constant shift to the right in the complex plane of all the thermal eigenmodes can be seen. Beside the difference in shape compared to \cref{Fig: flow_nonadia_thermal}, the growth rate is also slightly altered. The most unstable modes have become more unstable. From \cref{Eq: Ctcont} we have learned that the inclusion of a background flow does not alter the stability of the modes. However, including an azimuthal velocity or altering it does change other background parameters via the mechanical equilibrium that needs to be maintained. It manifests itself as a change in the pressure gradient or the magnetic field, see \cref{Eq: Mechequi}. Those quantities do influence the growth rate of the thermal continuum. Modifying the profile in an already flow-included equilibrium does only slighty change the growth rate. There is no change in growth rate when the axial velocity is altered. 

The second profile we consider here is a square root dependence on radius given by 
\begin{equation}
    v_{0\theta} = v_{0\theta b}\sqrt{r}.
\end{equation}

We take the parameter $\alpha$ as 0.5. This is a Keplerian rotation profile, in a different physical setting. The shape of the Doppler shifted thermal continuum shown in the right panel of \cref{Fig: flow_nonadia_thermal_alpha} is again different. The growth rates have increased, becoming even more unstable. The most unstable modes are still the outermost, radially. Due to the different Doppler shift they are now the least shifted along the real axis. 

\begin{figure}[htbp]
    \centering
    \resizebox{\hsize}{!}{\includegraphics{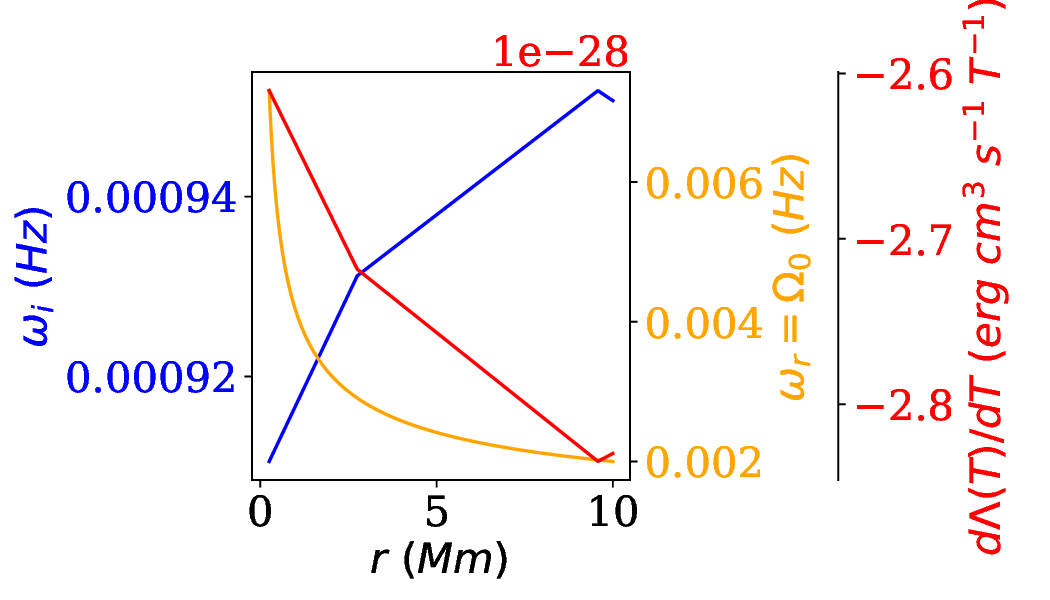}}
    \caption{The imaginary and real parts of the thermal continuum of the $\alpha = 0.5$ case. They are shown as blue and orange curves, respectively. This figure explains how the precise shape of the continuum relates to the temperature derivative of the cooling curve, denoted in red, used in this case.}
    \label{Fig: flow_nonadia_thermal_alpha05_split}
  \end{figure}

Compared to the spectra for the previous velocity profiles, this spectrum has some kinks in it. There is one at the upper-left part of the continuum and one around a growth rate of 0.00093~Hz. To see whether the Doppler shift, the imaginary part of the thermal continuum or the interplay of both is responsible, we look at the real and imaginary components of the thermal continuum. In \cref{Fig: flow_nonadia_thermal_alpha05_split} the real and imaginary parts of the continuum are shown with respect to the radial coordinate in orange and blue, respectively. We also plot the value of the derivative of the cooling rate for each radius in red. The Doppler shift is as expected from a square root azimuthal velocity profile. It is the imaginary part of the continuum that has the kinks. One can easily see that the kinks correspond to cusps in the derivative of the cooling rate with respect to the temperature. The cooling curve or its derivative do not need to be, and typically are not, smooth functions. This may even lead to discontinuous slow and thermal continua, as for example the slow continua for a solar coronal slab obtained by \citetads{2021SoPh..296..143C}. The different velocity profile in combination with the constraint of mechanical equilibrium translates to a slightly different temperature profile compared to the benchmark case. However, a different temperature profile also alters the cooling rate at every radial position and the derivative of the cooling rate. Hence, the change of such a small parameter may lead to different thermal and slow continua.

In all cases it is confirmed that the analytic expression matches the numerical results obtained using \textit{Legolas}. This is to be expected since the equations are derived from the general MHD equations in a cylindrical coordinate system using a generic background flow profile.

\subsection{Optically thin cooling curves}\label{Sec: TIflowccs}

\begin{figure}[htbp]
    \centering
    \resizebox{\hsize}{!}{\includegraphics{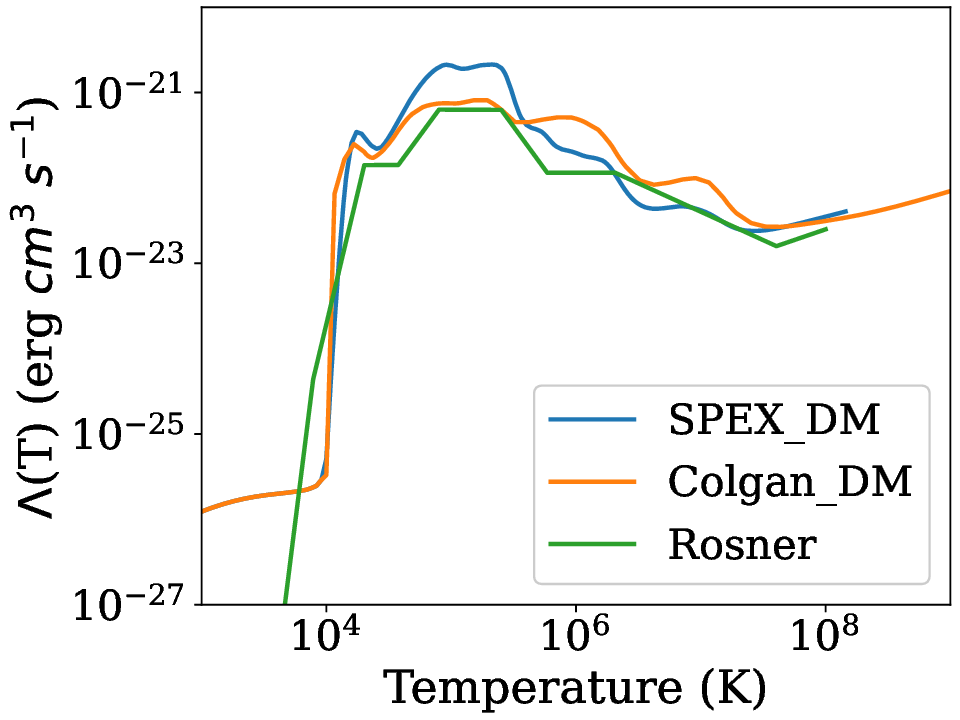}}
    \caption{The three optically thin cooling curves used in this work. The cooling rate is plotted in function of the temperature. The blue, orange, and green curves represent the SPEX\_DM, Colgan\_DM, and Rosner cooling curves, respectively.}
    \label{Fig: ccs}
\end{figure}

\begin{figure}[htbp]
    \centering
    \resizebox{\hsize}{!}{\includegraphics{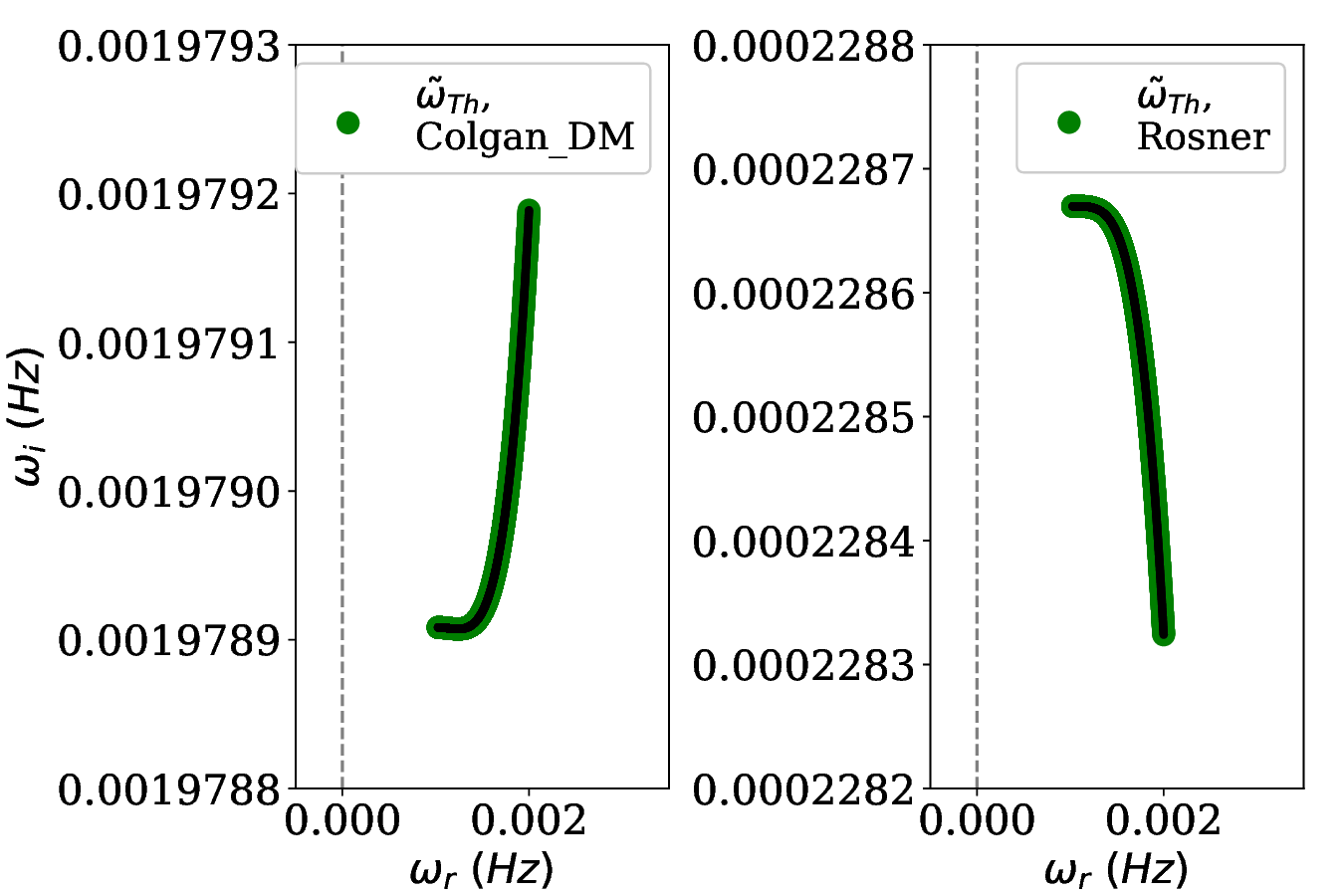}}
    \caption{The left and right panels shows the thermal continuum for the Colgan\_DM and Rosner cases, respectively. The analytic expressions are shown in green, while the \textit{Legolas} results are the black dots.}
    \label{Fig: flow_nonadia_thermal_ccs}
\end{figure}

In the benchmark case we used the SPEX\_DM cooling curve \citepads{2009A&A...508..751S}, however a wide variety of them is used in the literature. A similar cooling curve suitable for MHD simulations, being a more modern interpolated table, is the Colgan\_DM curve \citepads{2008ApJ...689..585C}. The curves are discussed in more detail in \citetads{2021A&A...655A..36H}. They are plotted in \cref{Fig: ccs}. The corresponding thermal continuum of the equilibrium, but with the Colgan\_DM cooling curve used instead, is shown in the left panel of \cref{Fig: flow_nonadia_thermal_ccs}. The only notable difference is the difference of factor two in the growth rate of the modes. The fact that growth and damping rates can differ due to the use of a different cooling curve has also been discussed by \citetads{2012A&A...540A...7S} and \citetads{2021A&A...655A..36H}. The outermost modes are still the most unstable.

A second interesting cooling curve is the Rosner curve \citepads{1978ApJ...220..643R,1982soma.book.....P}. It is also shown in \cref{Fig: ccs}. It is an older piece-wise power law with a lower radiative cooling rate at 1 million Kelvin, leading to expected longer growth and damping rates. In the right panel of \cref{Fig: flow_nonadia_thermal_ccs} the thermal continuum using the Rosner curve is shown. The growth rate is indeed significantly lower. However, what is most striking is the shape of the Doppler shifted continuum. It is mirrored with respect to the benchmark and Colgan\_DM cases. This is due to the fact that the outermost modes are not the most unstable. The innermost modes are the most unstable when using the Rosner curve because of the physical parameters of the background equilibrium. The innermost modes are shifted the least, creating the flipped shape of the Doppler shifted continuum. It is important to note that the choice of cooling curve can thus alter the location of the most unstable modes. The location of the most unstable modes can be determined from the spectrum, if the velocity profiles are known.

\subsection{Thermal conduction and axial wavenumber}

A physical effect that we did not yet take into account in the numerical investigation of the previous subsections is thermal conduction. Its influence has been extensively studied in the literature for static equilibria and in multidimensional simulations \citepads{1965ApJ...142..531F,1990ApJ...358..375B,2010ApJ...720..652S,2020A&A...636A.112C,2021MNRAS.505.5238J}. Thermal conduction smooths out temperature gradients and small perturbations with wavelengths shorter than a characteristic length scale. The thermal and slow modes might be damped depending on physical parameters of the equilibrium and the wavenumber. The Alfv\'en modes are not influenced. We restrict ourselves to field-aligned, parallel conduction.

\begin{figure}[htbp]
    \centering
    \resizebox{\hsize}{!}{\includegraphics{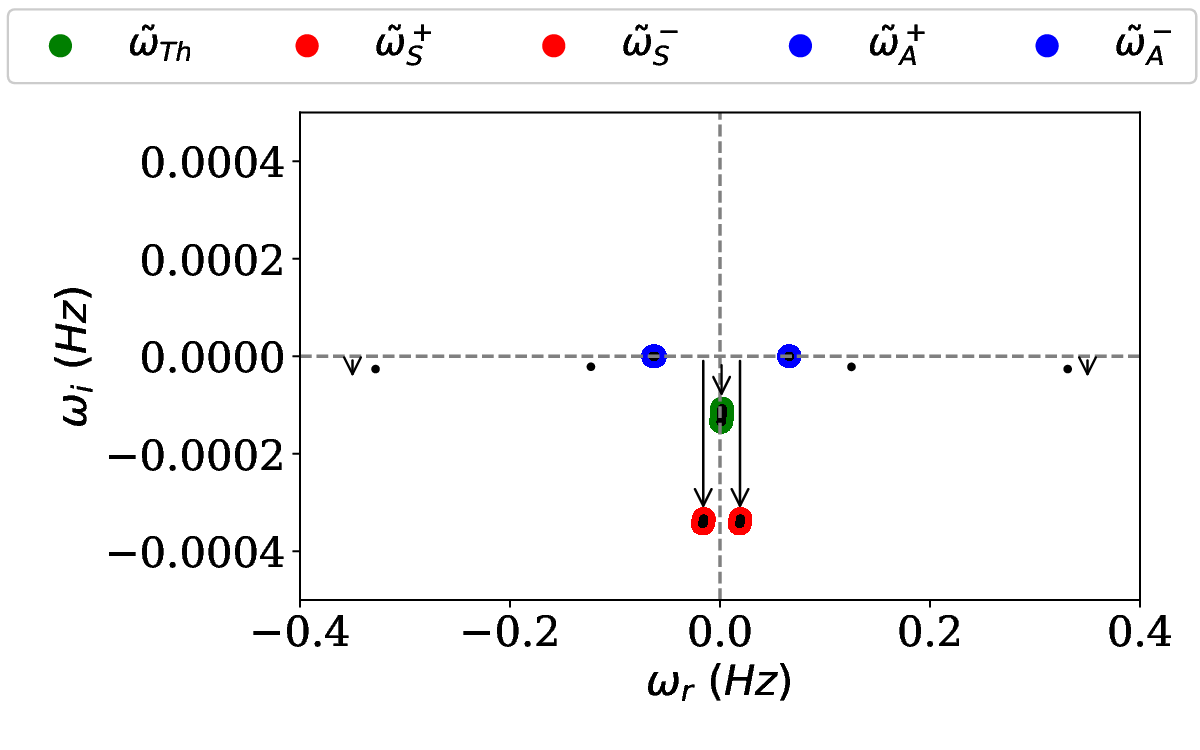}}
    \caption{The spectrum of the benchmark equilibrium with parallel thermal conduction included. The black dots are the results obtained with \textit{Legolas}. The green, red, and blue dots are the thermal, slow and Alfv\'en continua, respectively.}
    \label{Fig: flow_nonadia_cond_overview}
\end{figure}

In \cref{Fig: flow_nonadia_cond_overview} the spectrum is shown for the benchmark case, but with parallel conduction. The first thing to note is that all the continua obtained with \textit{Legolas} still confirm the analytic expressions. Secondly, the thermal continuum is now damped. The modes do not lead to thermal instability in this case. However, the shape in the complex plane is still the same. The slow continuum modes are more damped and the Alfv\'en modes are unaltered.

\begin{figure}[htbp]
    \centering
    \resizebox{\hsize}{!}{\includegraphics{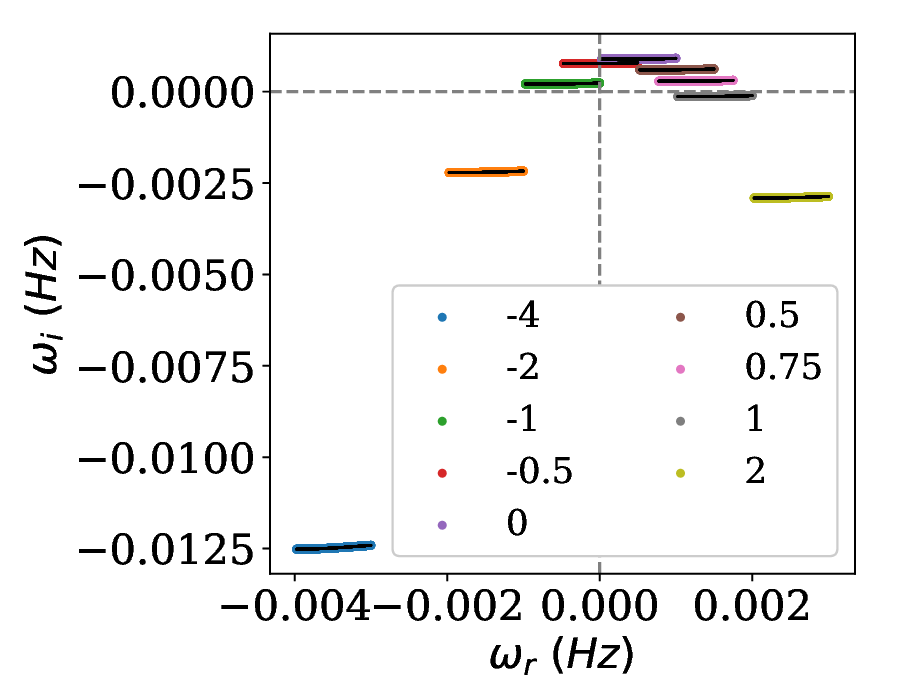}}
    \caption{The thermal continua for the different wavenumbers given in the legend in the complex plane. The black dots correspond to a run with a given value of $k$, as denoted by the colour of the overlapping thermal continuum.}
    \label{Fig: flow_nonadia_cond_dif_k}
\end{figure}

The previous results show that the plasma is not thermally unstable for a dimensionless axial wavenumber of unity. However, this purely coronal volume of plasma might still be unstable for different values of this wavenumber. The magnitude of the axial wavenumber can be related to the wavelength of the observed tornado structures. In the literature several values are quoted, see e.g. \citetads{2013ApJ...774..123W} and \citetads{2023SSRv..219....1T}. We take typical heights to be between 5 and 100 Mm. The associated dimensionless axial wavenumbers can be calculated as follows 
\begin{equation}
    \abs{k} = \frac{2\pi}{\lambda} L_u = \frac{2\pi \cdot 10^{\bm{9}}~cm}{[5 - 100]~Mm} = \frac{2\pi \cdot 10^9~cm}{[5 - 100] \cdot 10^8~cm} \approx [0.5 - 12].
\end{equation}
Shorter perturbations with larger axial wavenumbers might represent finestructure. In \cref{Fig: flow_nonadia_cond_dif_k} the thermal continua for different values of $k$ are shown in the complex plane. We varied $k$ from -4 to 2, where the minus-sign denotes the direction of the wavevector. We keep the azimuthal wavenumber fixed at 1. For every continuum the \textit{Legolas} results in black are accompanied by the analytic expression in colour. The continuum modes with an axial wavenumber of 0 are the most unstable. They are shifted to the right by a completely positive Doppler shift. Increasing the axial wavenumber dampens the thermal modes. The continua are also more shifted to the right. If the axial wavenumber is decreased, becoming more and more negative, the continua are more damped and shifted to the left. For small, negative axial wavenumbers the shift might not be completely positive, leading to a part of the real thermal continuum that is positive and one that is negative. Some modes of the continuum will be forward propagating, while others will be backward propagating.

\begin{figure}[htbp]
    \centering
    \resizebox{\hsize}{!}{\includegraphics{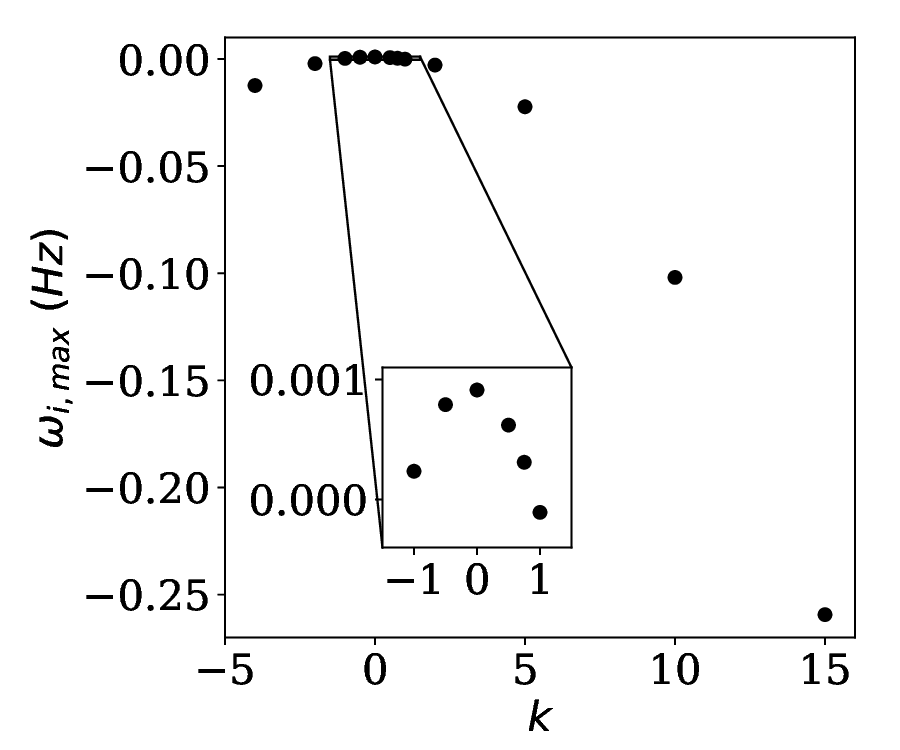}}
    \caption{The largest growth rate of the thermal continuum in function of the axial wavenumber. An inset is shown to zoom-in on the unstable wavenumber range.}
    \label{Fig: flow_nonadia_cond_maxomegai_k}
\end{figure}

In \cref{Fig: flow_nonadia_cond_maxomegai_k} we show the largest growth rate of the thermal continuum with respect to the axial wavenumber. This mode triggers and determines the thermal instability, if unstable. It can be seen that the growth rate is diminished for large $k$. Hence small perturbations are smoothed out. Perturbations with wavenumbers less than unity can unstable for this coronal volume of plasma, as can be seen in the inset. This corresponds to perturbations with the size of roughly 60 Mm. Note that the growth rates are different for negative and positive versions with the same $k$. This is because $k$ also modifies the thermal continuum through $F$ in \cref{Eq: Ctcont}.

Parallel thermal conduction has the same effect of damping the thermal modes, based on a cut-off wavelength. Small perturbations are damped for the equilibrium parameters considered here. However, a slight change to the temperature or density might be enough to make the medium unstable for certain wavenumbers which were previously damped.

\section{Discrete modes}\label{Sec: discretemodes}

Besides the continua, also discrete solutions to the linearised equations exist. While continuum modes have very localised singular eigenfunctions, the eigenfunctions of discrete modes vary with radius. Discrete modes exist in the form of well-known instabilities, e.g. in linear resistive MHD one frequently invokes the tearing mode \citepads{1963PhFl....6..459F}. However, discrete modes can also appear as a series of modes that cluster towards the edges of continua \citepads{1974PhFl...17..908G,1984PhyD...12..107G}. Stability of such modes depends on the physical parameters of the equilibrium. This has been studied for Alfv\'en and slow modes by \citetads{1984PhyD...12..107G}. \citetads{1991SoPh..134..247V} proved that discrete thermal modes can also exist and argued that they can alter the global, thermal stability of an equilibrium. The influence of a background flow on adiabatic discrete modes was studied by \citetads{2004JPlPh..70..651W}.   

In this section, we extend on the literature of discrete modes by including both a background flow and non-adiabatic effects. We look in particular at the discrete thermal and slow modes. In the first subsection we derive an analytic approximation for the frequency of the discrete modes. In the following subsections we investigate the discrete modes numerically using \textit{Legolas} and visualise them in 3D.

\subsection{Analytic investigation}

In this section we perform a WKB analysis to derive an approximate dispersion relation near an internal extremum of the thermal or slow continuum. Discrete modes might cluster towards such a point. We follow the methodology of \citetads{1984PhyD...12..107G}, \citetads{1991SoPh..134..247V}, and \citetads{1993SoPh..144..267K}.

The starting point is the second-order differential equation \cref{Eq: 2orderdiff}, which can be written in the form 
\begin{equation}\label{Eq: 2orderdiff_disc}
    D\left[ f(r) D(r\xi_r) \right] - g(r) r\xi_r = 0
\end{equation}

\noindent with the coefficients $f(r)$ and $g(r)$ dependent on the coefficients $C_0$ to $C_t$ given by \crefrange{Eq: C0}{Eq: C_t}. A fundamental asssumption of the WKB approximation is that the coefficients $f(r)$ and $g(r)$ are only weakly varying. We then assume a solution of the form
\begin{equation}\label{Eq: WKB sol}
    r\xi_r = p(r) e^{i\int q(r) dr}
\end{equation}

\noindent where $p(r)$ is the amplitude modulation function and $q(r)$ is the local radial wavenumber \citepads{1993SoPh..144..267K}. The exponential part is assumed to rapidly vary compared to the prefactor $p(r)$ and with respect to the characteristic length scale of the equilibrium $L$, i.e. $q^2 L^2 \gg 1$. The applicability of the WKB results is thus limited to modes of high radial order. The expressions for $p(r)$ and $q(r)$ can be determined by substituting \cref{Eq: WKB sol} into \cref{Eq: 2orderdiff_disc} and only considering the leading order terms in the inhomogeneity. We obtain
\begin{align}
    p &\approx (-fg)^{-1/4},\\
    q &\approx (-g/f)^{1/2}.
\end{align}

The latter relation defines a local dispersion relation that connects the local radial wavenumber $q$ to the frequency. Expressing $f$ and $g$ in terms of the coefficients of the first-order differential equations, \crefrange{Eq: C0}{Eq: C_t}, yields
\begin{equation}
    q^2 = -\frac{g}{f} = -\frac{C_2}{C_0} \left[ \frac{C_1^2 + C_2C_3}{C_0C_2} + D\left[ \frac{C_1}{C_2}\right]  \right] 
\end{equation}

Multiplying both sides with $C_0$ and substituting its expression, \cref{Eq: C0}, gives
\begin{equation}\label{Eq: Localdisp}
    r[\rho_0\tilde{\omega}^2 - F^2] C_tq^2 =  -C_2\left[ \frac{C_1^2 + C_2C_3}{C_0C_2} + D\left[ \frac{C_1}{C_2}\right]  \right] 
\end{equation}

We now determine approximate solutions to this local dispersion relation near the extremum of the slow and thermal continua. To that end we suppose that the continua are sufficiently far apart in the complex plane. The slow and thermal continua are strongly connected because of their coupling due to the non-adiabatic, radiative effects. Both kind of modes are solutions of the same equation \cref{Eq: Ctcont}. We consider a Doppler-shifted thermal or slow continuum that has an internal extremum at $r_0$, such that $\tilde{\omega}_0 = \tilde{\omega}(r_0)$ and $C_t(r_0, \tilde{\omega}_0)=0$. Such a continuum can be found or constructed using the equilibrium profiles.

We expand the coefficient $C_t$ in the neighbourhood of the internal extremum
\begin{equation}
    C_t(r_0, \tilde{\omega}_0 + \epsilon) \approx C_t(r_0, \tilde{\omega}_0) + \pdv{C_t}{\tilde{\omega}}\epsilon = \epsilon\pdv{C_t}{\tilde{\omega}},
\end{equation}

\noindent where $\epsilon$ is a small, complex perturbation. The left-hand side of the local dispersion relation becomes
\begin{equation}
    r[\rho_0\tilde{\omega}_0^2 - F^2] \pdv{C_t}{\tilde{\omega}} \epsilon q^2, 
\end{equation}

\noindent where every equilibrium quantity is evaluated at $r=r_0$. The left-hand side is thus accurate up to first-order of $\epsilon$. The right-hand side of the local dispersion relation is expanded to zeroth-order of $\epsilon$ and evaluated at $r=r_0$. After a lengthy derivation $\epsilon$ can be determined. The approximation of the eigenfrequency is finally given by
\begin{equation}\label{Eq: omegaapprox}
    \tilde{\omega} \approx \tilde{\omega}_0 + \epsilon = \tilde{\omega}_0 + \frac{1}{q^2}\zeta,
\end{equation}

\noindent with $\tilde{\omega}_0$ the continuum frequency around which was expanded. The parameter $\zeta$ is given by
\begin{equation}\label{Eq: zeta}
    \begin{split}
    \zeta = & \frac{i\left[\tilde{\omega}_0\rho_0 + i(\gamma - 1)\left[\frac{\kappa_\parallel}{B_0^2}F^2 + \rho_0 \mathcal{L}_T\right]\right]}{[\rho_0\tilde{\omega}_0^2 - F^2] \pdv{C_t}{\tilde{\omega}}}   \Biggl\{   \rho_0^2\tilde{\omega}_0^4 [\rho_0\tilde{\omega}_0^2 - F^2]\\
    & - r\rho_0^2\tilde{\omega}_0^4 D\left[ \frac{B_{0\theta}^2 - \rho_0 v_{0\theta}^2}{r^2} \right] -\frac{4\rho_0\tilde{\omega}_0^2}{r^2} \frac{[FB_{0\theta}+v_{0\theta}\rho_0\tilde{\omega}_0]^2}{\rho_0\tilde{\omega}_0^2 - F^2} \\
    &\cross [k^2B_0^2 + \rho_0\tilde{\omega}_0^2 - F^2] + \frac{4m\rho_0\tilde{\omega}_0^2}{r^3} [FB_{0\theta}+v_{0\theta}\rho_0\tilde{\omega}_0]\\
    &\cross [2B_{0\theta}^2 - \rho_0v_{0\theta}^2] -\frac{1}{r^2} \left(\frac{m^2}{r^2} + k^2\right) \left[   [2B_{0\theta}^2 - \rho_0 v_{0\theta}^2]^2 \rho_0 \tilde{\omega}_0^2 \right.\\
    &\left. - \rho_0^2 v_{0\theta}^2 [2B_{0\theta}\tilde{\omega}_0 + v_{0\theta}F]^2          \right]  - r\rho_0^2\tilde{\omega}_0^2  \left[D\left[\frac{v_{0\theta}^2F^2}{r^2}\right] \right.\\
    &\left. + 2 \tilde{\omega}_0 D\left[\frac{v_{0\theta}B_{0\theta}F}{r^2}\right] + \tilde{\omega}_0^2 D\left[\frac{\rho_0 v_{0\theta}^2}{r^2}\right]    \right]   \Biggr\}
    \end{split}
\end{equation}
\noindent with 
\begin{equation}
    \begin{split}
    \pdv{C_t}{\tilde{\omega}} =& 2 i \rho_{0}^2 \tilde{\omega}_0^2[\gamma p_{0} + B_0^2] - \frac{p_{0} (\gamma - 1)}{ \tilde{\omega}_0} \left[\frac{\kappa_\parallel}{B_0^2}F^2 + \rho_0 \mathcal{L}_T - \frac{\rho_{0}^2 \mathcal{L}_\rho}{T_0} \right]\\
     & \cross  \left[ \rho_{0} \tilde{\omega}_0^2  + F^2 \right] - \rho_{0} \tilde{\omega}_0 (\gamma - 1) B_0^2 \left[  \frac{\kappa_\parallel}{B_0^2}F^2 + \rho_0 \mathcal{L}_T\right].
    \end{split}
\end{equation}
\noindent All parameters are evaluated at the local extremum of the continuum, ($r_0$,$\tilde{\omega}_0$). Furthermore, we used the fact that $D(C_t)(r_0, \tilde{\omega}_0) = 0$ to simplify the derivation.

When background flow is neglected, our approximated solution reduces to equations 4.10 and 4.11 of \citetads{1991SoPh..134..247V}, taking into account the difference in notation. Moreover, in the case of an ideal non-adiabatic plasma, the expressions reduces to those for discrete slow modes found by \citetads{1984PhyD...12..107G}. A Frobenius expansion around the extremum would give similar results. \citetads{2004JPlPh..70..651W} derived a local clustering criterium for discrete slow modes using this method. Our expression reduces to their result. Hence, the results obtained in this chapter reduce to all simpler cases. 

\cref{Eq: omegaapprox} is not a closed instability criterion. Nonetheless, it can be used to predict the appearance of discrete modes numerically without solving the differential equations, as we show in \cref{Sec: disc_num}. The parameter $\zeta$, given by \cref{Eq: zeta}, contains many free parameters and profiles. This makes the expression quite unclear and hard to handle. With respect to the background flow, it is important to note that the axial component does not play a role. The azimuthal component is nevertheless omnipresent in the expression. 

The sign of $\zeta$ is crucial in the determination of the thermal stability of an equilibrium. An equilibrium is susceptible to thermal instability when thermal or slow modes are unstable, i.e. have a positive growth rate. The most unstable mode determines the evolution of the system. Discrete modes can alter the stability when they are more unstable. Discrete modes can exist when they lie above a maximum or below a minimum of the continuum. Otherwise, they would be absorbed into it. In the case of a maximum in the continuum $\zeta$ needs to be positive, vice versa for a minimum. An interesting case would be an equilibrium where all the continua are damped, but with unstable discrete thermal modes. The condensations of equilibria with the most unstable modes being discrete modes rather than continuum modes also have a different shape, a spread-out profile instead of a very localised perturbation. 

Note that a similar expression can be derived for the non-adiabatic Doppler shifted Alfv\'en continuum. However, we did not pursue it here because it deviates too far from our main topic of thermal modes. Discrete Alfv\'en modes were studied by \citetads{1984PhyD...12..107G} in an ideal plasma. \citetads{1991SoPh..134..247V} and \citetads{1993SoPh..144..267K} included radiative cooling and thermal conduction. The adiabatic case with a background flow was studied by \citetads{2004JPlPh..70..651W}.

\subsection{Numerical investigation}\label{Sec: disc_num}

The spectra shown in \cref{Sec: num_inv_cont} do not show any discrete thermal modes. This is because of the equilibrium parameters used. The thermal continuum does not have extrema for modes to cluster to. In this section we look at a few equilibria that do have an extremum in their thermal continuum. We verify our expression to predict the existence of discrete thermal modes derived in the previous subsection.

\subsubsection{Setup}\label{Sec: disc_set}

\begin{figure}[htbp]
    \resizebox{\hsize}{!}{\includegraphics{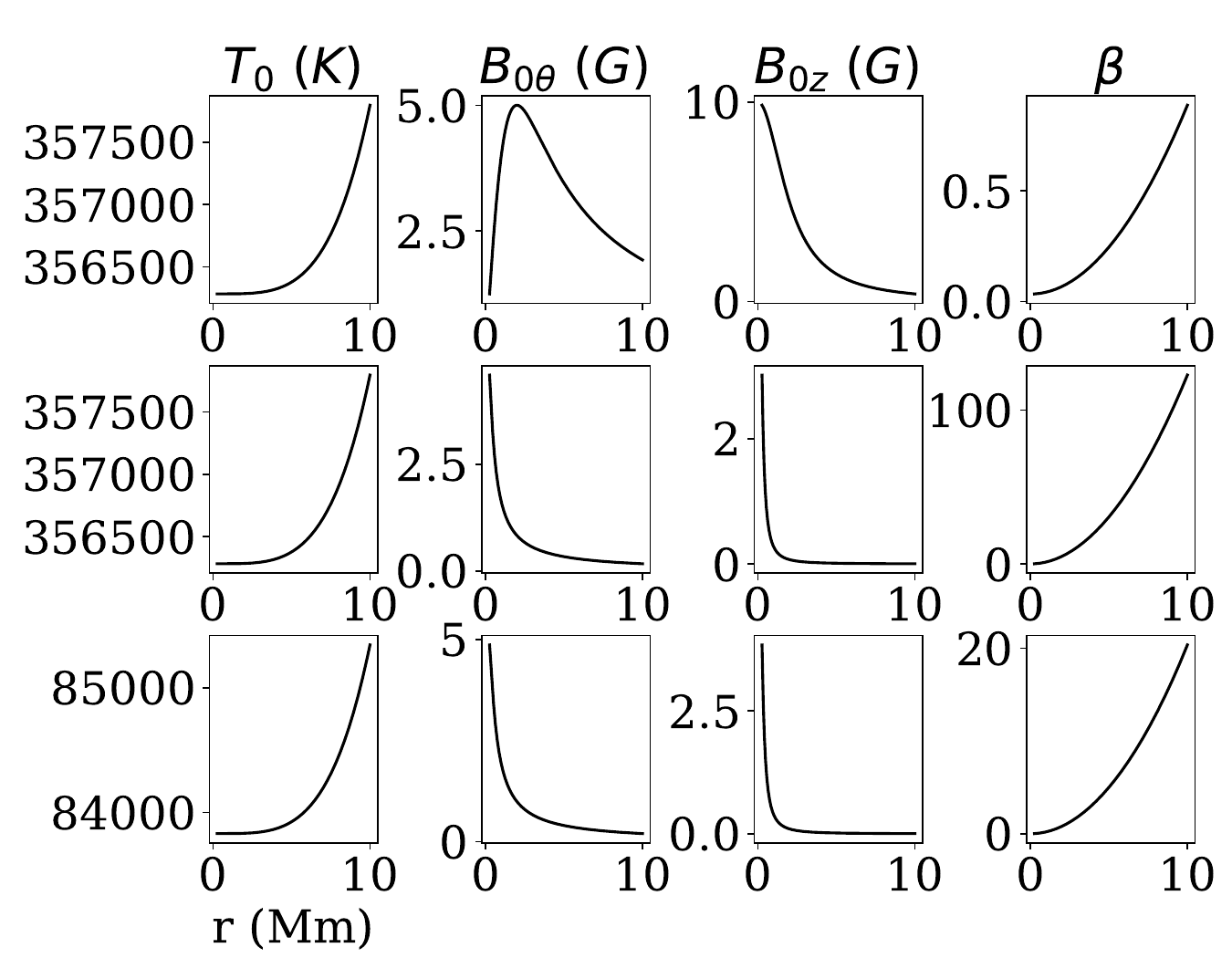}}
    \caption{The background temperature, azimuthal magnetic field, axial magnetic field, and plasma beta profiles of the three equilibria with discrete thermal modes. Each row corresponds to a different case.}
    \label{Fig: equilibria_discreetflow}
\end{figure}

We base the three equilibria on the benchmark setup discussed in \cref{Sec: Setup_TIflow}. We keep most parameters the same. The wavenumber $k$ is for all three cases set to 0.5. The two parameters that are changed are the temperature at the axis and the magnetic inverse pitch, $\mu$. Varying the inverse pitch affects the shape of the magnetic field components. In \cref{Fig: equilibria_discreetflow} the profiles for the background temperature, azimuthal and axial magnetic field components, and the plasma beta are shown. Each row corresponds to a case. The temperature of case 1 and 2 are the same and around $357000$~K, which is about a third of the benchmark case. The optically thin cooling rate is different, as is its derivative. This influences the thermal continuum a lot and can cause extrema. The difference between the first and second case is the inverse pitch, this is set to 5 for the former and to 60 for the latter. The field is predominantly vertical for low inverse pitch. The plasma beta also remains below unity for the whole domain, as can be seen in the right-most panel of the top row of \cref{Fig: equilibria_discreetflow}. For case two the magnetic field drops off quickly, i.e. is more centralised around the axis. In the third case the inverse pitch is set to 50 and the temperature is again lowered. The derivative of the cooling rate with respect to temperature is positive for low temperatures, around $84000$~K for the third case. We therefore expect a damped thermal continuum.

\subsubsection{Results}

\begin{figure}[htb!]
    \resizebox{\hsize}{!}{\includegraphics{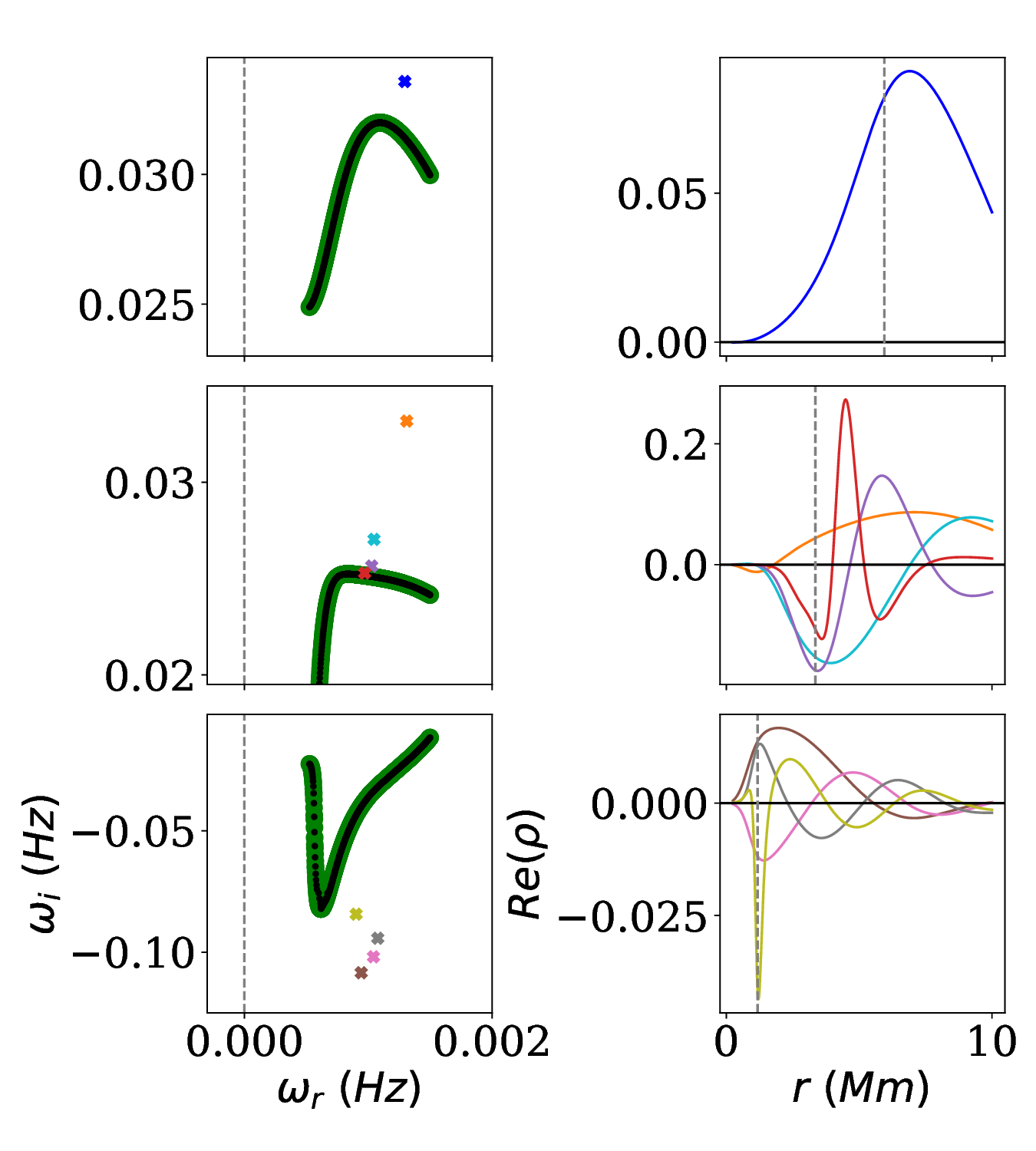}}
    \caption{The thermal continua, discrete thermal modes, and corresponding density eigenfunctions for the three cases. Each row corresponds to a different case. The \textit{Legolas} results are shown as black dots. The analytic thermal continuum is overlaid in green. The discrete thermal modes are marked by coloured crosses. The colour of the eigenfunctions match the colour of the discrete modes. Grey dotted lines in the right panels denotes the radial location of the extremum of the thermal continuum.}
    \label{Fig: spectr_disc_th}
\end{figure}

The results of the \textit{Legolas} runs for the three equilibria described previously are shown in \cref{Fig: spectr_disc_th}. The left panels contain the thermal continua, with the discrete thermal modes marked, in the complex plane for the three cases. For the second case only a part of the continuum is plotted. This is to focus on the discrete modes. The right panels show the density eigenfunctions of the marked discrete modes. A grey dotted line denotes the radial location of the extremum. Each row of \cref{Fig: spectr_disc_th} corresponds to one of the three equilibria. 

First of all, in contrast to the thermal continua discussed in \cref{Sec: num_inv_cont}, there is an extremum. In the first and second case it is a maximum. Hence, a different temperature and by extension cooling rate and its derivatives, has a large impact on the continuum. For the case for which the plasma beta is still completely below unity, case 1, there is one discrete mode. The eigenmode for the density is positive as to be expected for the growth of a condensation. For the second equilibrium, shown in the middle row of \cref{Fig: spectr_disc_th}, there are four discrete modes. The number of nodes increase as the frequency decreases in an anti-Sturmian fashion. The third equilibrium has a much lower temperature. The thermal continuum is completely damped due to the derivative of the cooling rate being positive. The continuum has a minimum and there are four discrete modes that cluster towards that minimum. This is a Sturmian sequence. 

For all three cases we can check whether our expression found by the WKB-approximation gives the expected sign. As there is still the unknown parameter $q$ in the expression of $\epsilon$, in \cref{Eq: omegaapprox}, only the sign of $\zeta$ can be used to predict the existence of discrete modes. The magnitude of $\zeta$ is not important because $q$ is unknown and assumed to be large, ensuring that $\epsilon$ is small. When there is a maximum in the thermal continuum for discrete modes to cluster to the imaginary part of $\zeta$ has to be positive, and vice versa for a minimum. The discrete modes are on the right side of the extremum for all three cases and hence the real part is expected to be positive. The calculated values are $0.022 + 0.09i$, $0.003 + 0.66i$, $0.448 - 10.42i$ for the first, second, and third case, respectively. The signs are all in perfect agreement with expectations. It should be noted that the precision in the numerical values of these numbers is very sensitive to round-off since quantities that are of very similar size are subtracted. Small differences due to normalisation and physical constants used also affects them. 

\subsection{Visualisation of eigenmodes}

By solving the eigenvalue problem for a given equilibrium, you obtain the frequencies and eigenfunctions of the eigenmodes of the system. \citetads{alma9993481308901488} describes how \textit{Legolas} solves the eigenvalue problem and calculates the eigenfunctions. For this to work, we assumed a wave mode representation of the perturbations. In \cref{Eq: Fourierdecomp}, this was given in cylindrical coordinates by 
\begin{equation}
    f_1 (\bm{r},t) = f'(r)e^{i(m\theta + k z-\omega t)},
\end{equation}

\noindent where the primed $f'(r)$ are the amplitudes of the waves, i.e. the eigenfunctions. A new feature in \textit{Legolas 2.0} uses this equation to visualise the modes in multiple dimensions, as presented in \citetads{2023arXiv230710145C}. Setting values for $t$, $\theta$, and/or $z$ allows one to calculate the perturbation at that given location using the eigenfunction, wavenumbers, and frequency. Depending on for which parameter a range is supplied, the temporal evolution, spatial slices or full 3D datacubes can be visualised. It is important to stress that the views obtained using this method represent the linear evolution of the waves and not the non-linear behaviour. Nevertheless, it is a useful and simple way to gain information without performing full non-linear multidimensional MHD simulation.

We here look at the discrete modes and use the eigenfunctions and frequencies determined in the previous subsection. All the quantities plotted in the figures of the subsection are dimensionless, while the axis are rescaled to represent the actual length and time scales. We will use a colourscale for which darker represents a larger density. 

\begin{figure}[htbp]
    \centering
    \resizebox{\hsize}{!}{\includegraphics{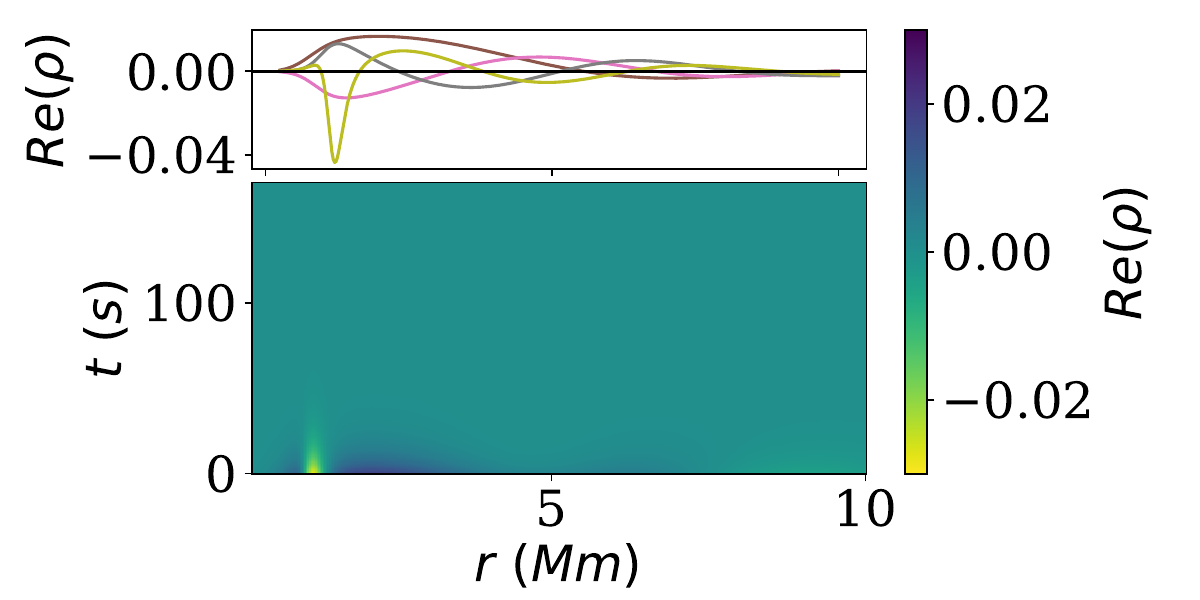}}
    \caption{The temporal view for discrete modes of the third case, the damped eigenmodes. The eigenfunctions are shown in the top panel and coloured according to \cref{Fig: spectr_disc_th}.}
    \label{Fig: temporal_damped}
\end{figure}

Let us first look at the temporal evolution of the least interesting case, the third equilibrium with damped discrete modes. The temporal evolution can be visualised by choosing a fixed $\theta$ and $z$ to look at and vary $t$. We took the default value of 0 for $\theta$ and $z$. We use a superposition of the four discrete modes with an amplitude ratio of unity for the four modes. The evolution is shown in the bottom panel of \cref{Fig: temporal_damped}. The initial perturbation based on the eigenfunctions is excited at $t=0$ and also depicted in the top panel. The eigenfunctions are coloured using the same colourscheme as in \cref{Fig: spectr_disc_th}. Since the growth rates are negative the modes are damped. This is clearly visible. Locations where the amplitude is bigger take more time to diminish.

\begin{figure}[htbp]
    \centering
    \resizebox{\hsize}{!}{\includegraphics{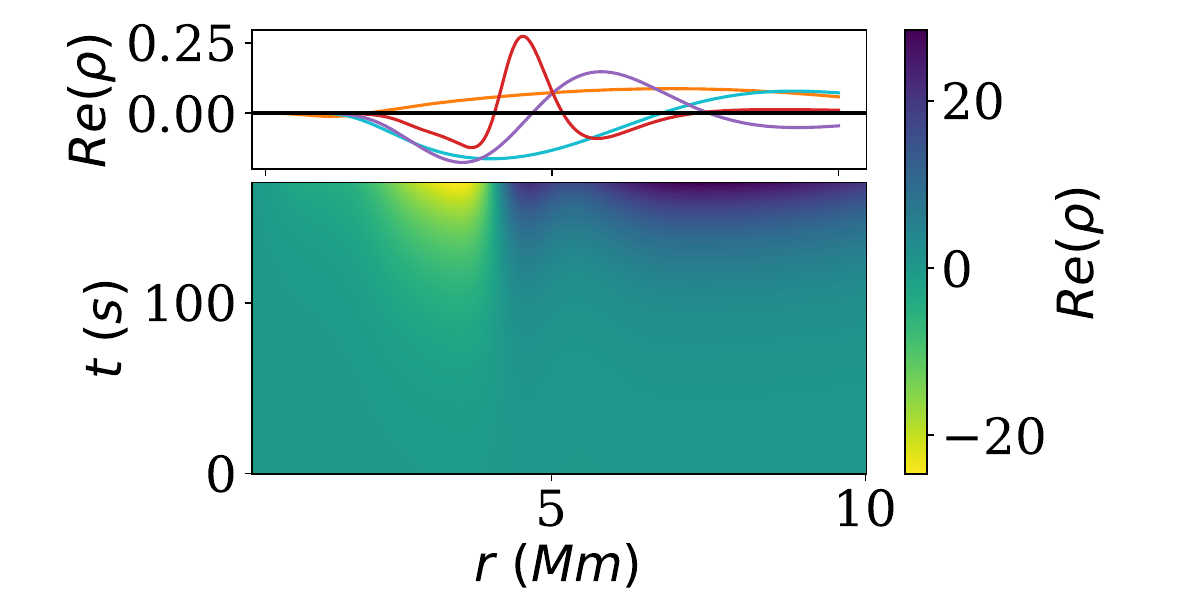}}
    \caption{The temporal view for discrete modes of the second case, the unstable eigenmodes. The eigenfunctions are shown in the top panel and coloured according to \cref{Fig: spectr_disc_th}.}
    \label{Fig: temporal_excited}
\end{figure}

The discrete modes of the second equilibrium, set up in \cref{Sec: disc_set}, are of more physical relevance. The four discrete modes have positive growth rates and are thus unstable. The temporal evolution is shown in \cref{Fig: temporal_excited}. Indeed, the modes become amplified. At later times it can be seen how the density is spread along the radius. The superposition of the eigenfunctions is negative between 20~Mm and 40~Mm, as can be seen in the top panel. This leads to a decrease in density in this region. In the outer part the density is increased. The discrete modes have a more global behaviour, in contrast to the ultra-localised eigenfunctions of the continuum modes. We did not favour any of the discrete modes over the others, i.e. none of the eigenfunctions were scaled differently.

\begin{figure}[htbp]
    \centering
    \resizebox{\hsize}{!}{\includegraphics{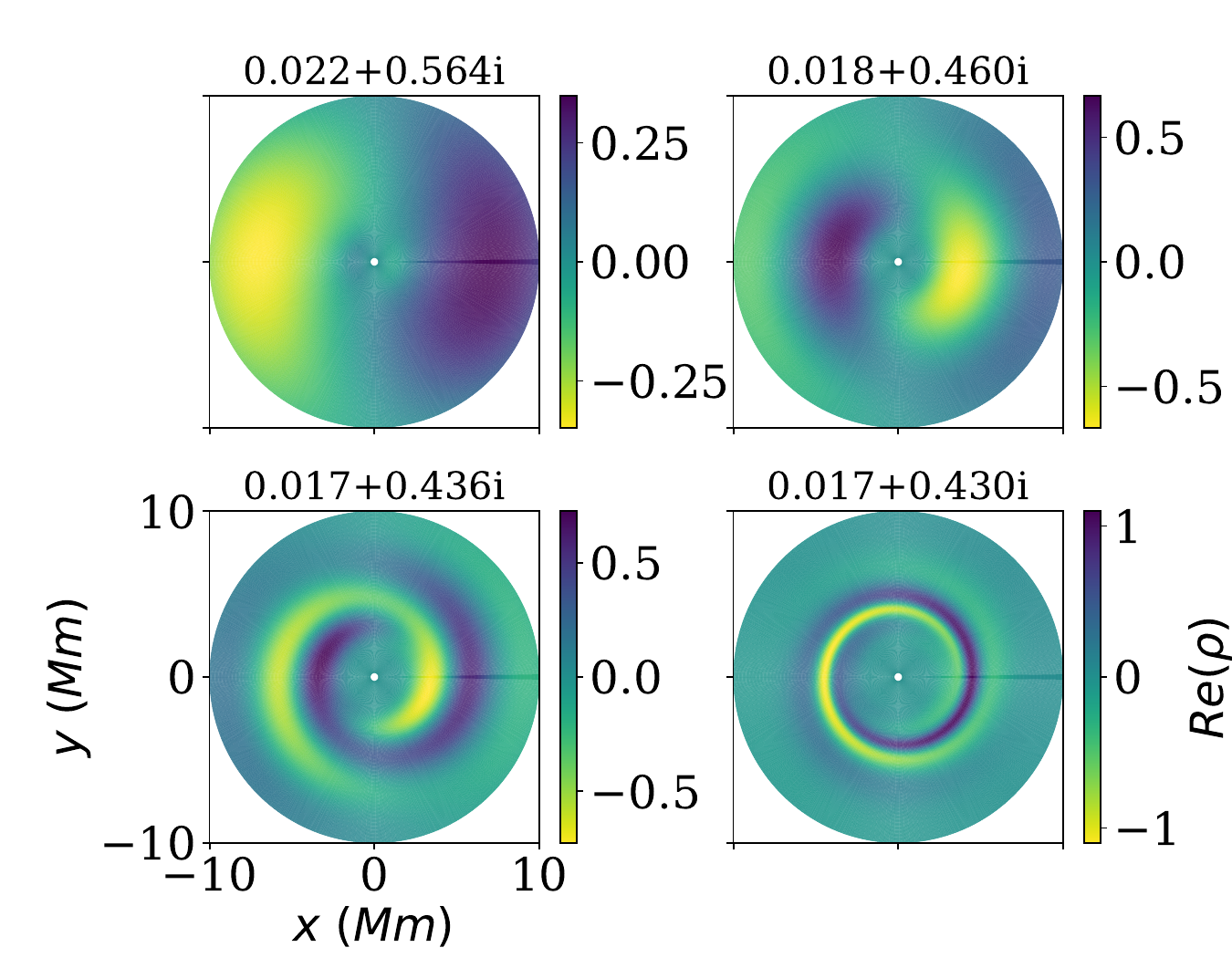}}
    \caption{Spatial cuts through the cylinder at $z=0$ and $t=0$, for the four discrete eigenmodes of the second equilibrium. From left to right, top to bottom each panel visualises a different mode starting from the mode with the least number of nodes to the one with the most nodes in the eigenfunction.}
    \label{Fig: spatial_theta-r}
\end{figure}

In \cref{Fig: spatial_theta-r} we present the 2D slice at constant height $z=0$ for the four discrete modes of the second equilibrium in the different panels for each mode. The top left panel contains the angular distribution of the most unstable mode, the mode with the largest growth rate and with the least nodes because of the Sturmian character of the sequence. The other modes are visualised in the other three panels, from left to right and top to bottom, the modes are ordered following the Sturmian sequence. We look at $t=0$, to negate the effect of the temporal evolution and to focus on the angular distribution.     

There is a clear difference between the left and right side of the slices. In the case of the most unstable mode in the top left panel, a condensation forms in the right while the left is vacated. This is due to the azimuthal wavenumber of the mode. The eigenmodes that we considered all have $m=1$ and thus are non-axisymmetric, like a `kink' mode. From a mathematical point of view this can be understood by looking at the angular part of \cref{Eq: Fourierdecomp}, $e^{im\theta}$.

For $m=0$ modes the exponential is constant and independent of the angle. However, for other values of $m$ this is not the case. For $m=1$ modes the exponential varies between -1 and 1, taking those values at $\theta=\pi$ and $\theta=0$ respectively. The magnitude of the eigenfunction is thus opposite on the other side of the axis. When going through the panels, it can also be seen that the modes become increasingly localised as they approach the continuum.

\begin{figure}[htbp]
    \centering
    \resizebox{\hsize}{!}{\includegraphics{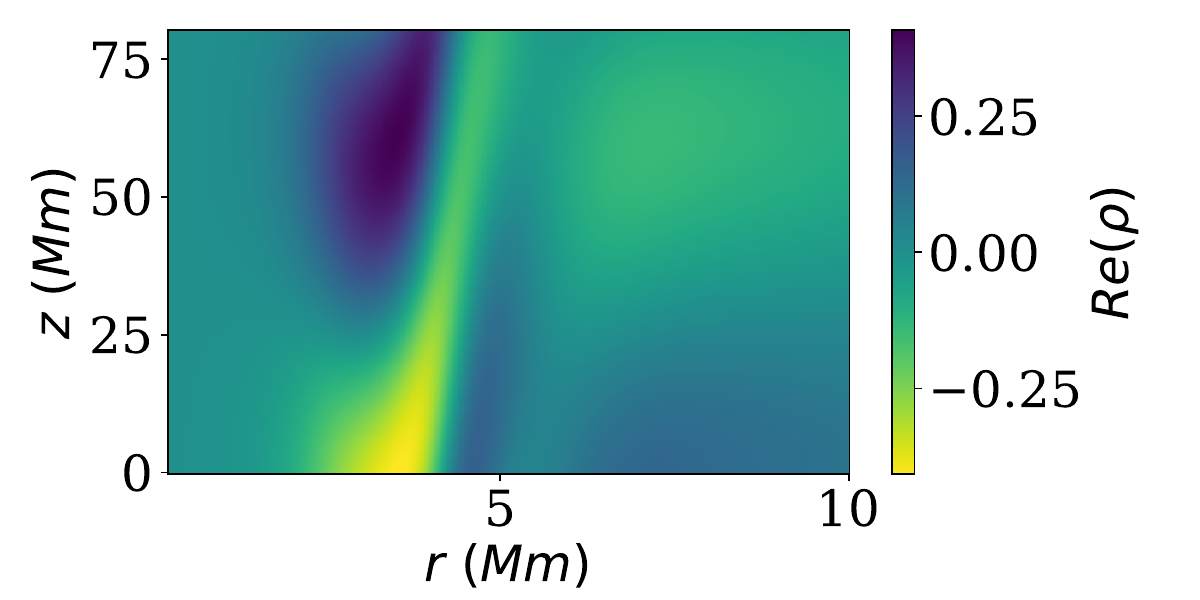}}
    \caption{The spatial cut at a fixed angle, the $z$-$r$ plane, for the superposition of the discrete modes of case 2 at time $t=0$.}
    \label{Fig: spatial_z-r}
\end{figure}

The last 2D slice that we present here is one at constant angle, $\theta=0$, and at the initial time. \cref{Fig: spatial_z-r} shows the variation of the density eigenfunction of the superposition of the four discrete modes of the second equilibrium with height for each radial position. The height is here varied from 0 to 8 Mm in the coordinate $z$. This corresponds the approximate height for prominence tornado structures. Due to the shape of the eigenfunctions the most dense region is between 40 and 80 Mm in height and 2 and 4 Mm in width. The condensations formed by these eigenmodes are thus more patchy in space.

\begin{figure*}[htbp]
    \centering
    \includegraphics[width=17cm]{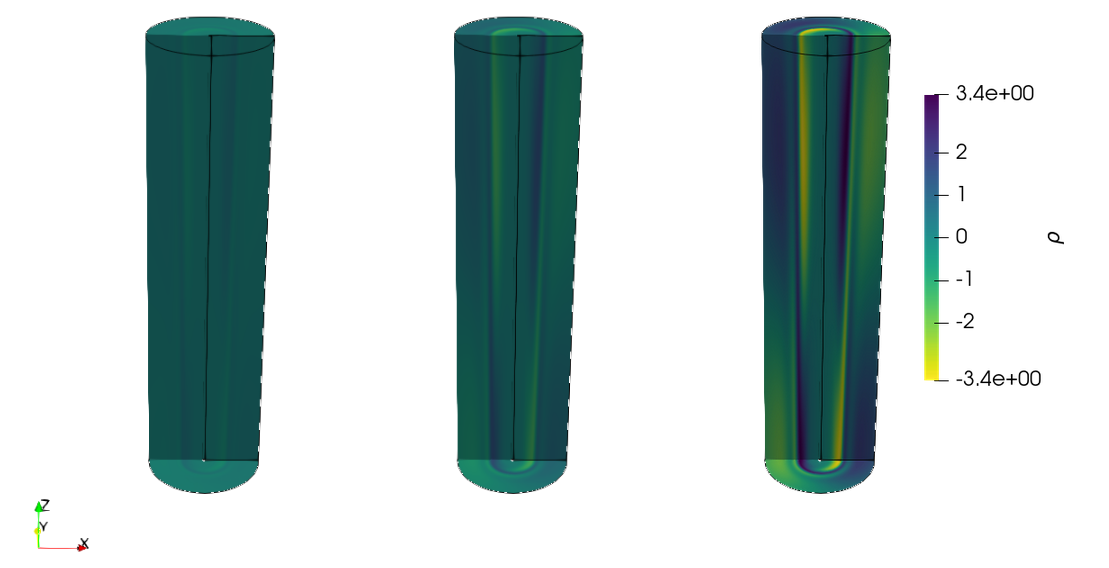}
    \caption{The 3D visualisations of the density perturbation throughout the cylinder at $t=0$, $t= 3$, and $t=5$.}
    \label{Fig: 3Ddensity}
\end{figure*}

The data cube can also be visualised in 3D. We exported the equilibrium and eigenfunctions for the four unstable discrete modes using the functionality of \textit{Legolas}. The obtained files were visualised using ParaView\footnote{\url{http://paraview.org}}. In \cref{Fig: 3Ddensity} we show the evolution of the density perturbation in the cylinder of height 80~Mm. Cuts are made for clarity. The left cylinder is at the initial time, the second at $t=3$, and the third at $t=5$. The density and noted timestamps are dimensionless. The density perturbation grows over time based on the initial shape of the eigenfunctions. The 2D views can be observed very clearly in the 3D images, by considering the cuts that have been made. The half of the bottom circle depicts the variation by the angle, as in \cref{Fig: spatial_theta-r}. The variation with height of \cref{Fig: spatial_z-r} is clearly seen in the right half of the open cylinder. The large imaginary part of the frequency of the superposition of the discrete modes dominates the temporal behaviour of the modes, whereas there appears little to no influence of the much smaller real part of the frequency.

\begin{figure}[htbp]
    \centering
    \resizebox{\hsize}{!}{\includegraphics{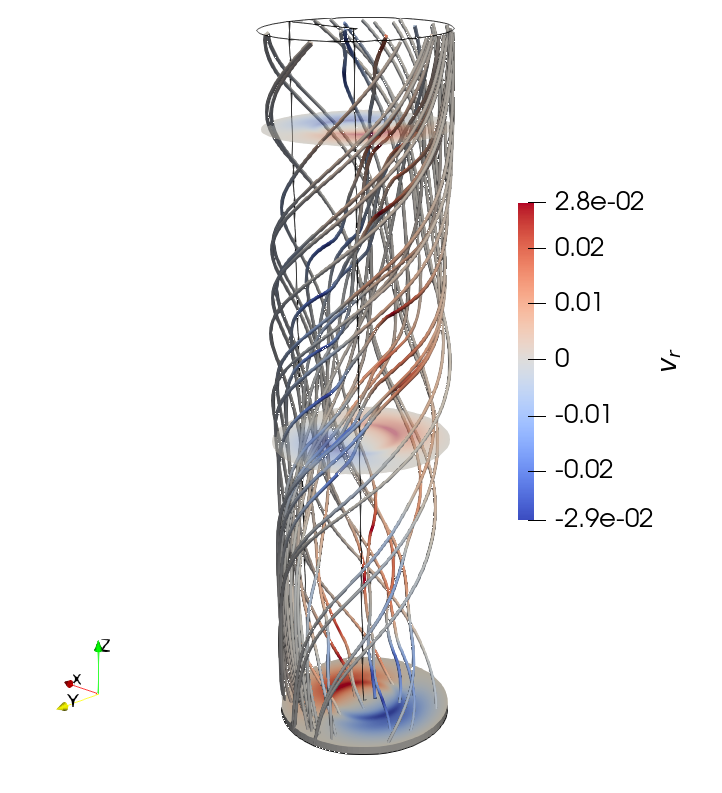}}
    \caption{The field lines of the total flow field. The field lines and slices are coloured according to the $v_r$ component of the velocity perturbation.}
    \label{Fig: 3Dflow}
\end{figure}

The streamlines of the total velocity field, background field plus perturbation, is shown in 3D in \cref{Fig: 3Dflow}. The field lines are coloured according to the local $v_r$ component of the velocity perturbation. There are three slices with $v_r$ at different heights provided. The top two are made a little bit more transparent in order to see the field lines more clearly. The total velocity field is mostly helical. However, due to the perturbation it is more complex than the equilibrium background flow, as shown in \cref{Fig: 3Dbench}. The added radial component distorts it. The radial component varies with height, as can be seen from the slices. The field lines can be seen moving inwards and outwards depending on their location. It should be noted that we did not rescale the eigenfunctions. Since they are of roughly the same order of magnitude as the background quantities, their influence is expected to be large.

\section{Discussion and conclusions}\label{Sec: sum-dis_TIflow} 

Flows are of the upmost importance in understanding the evolution of plasma in dynamical environments. Due to the increased complexity that they introduce, they are typically neglected when studying waves via MHD spectroscopy. Solar tornadoes are a class of prominences that appear to be rotating and could be formed by thermal instability. We therefore extended the work of \citetads{1991SoPh..134..247V} by including a helical background flow. 

We went through the linearisation and Fourier-analysis of the non-adiabatic MHD equations in a cylindrical coordinate system with a generic helical flow. The set of equations is transformed into two first-order differential equations, one for $v_r$ and one for $Y$, in the spirit of the work by \citetads{1974PhFl...17.1471A}. The equations differ from those derived by \citetads{1991SoPh..134..247V}. Most importantly, the frequency is Doppler shifted. Additional terms appear they are all dependent on the azimuthal background velocity $v_{0\theta}$. There are no terms that depend on the axial background velocity $v_{0z}$. This means that azimuthal flow is of the utmost importance, which was also noticed by \citetads{2004JPlPh..70..651W} for adiabatic equilibria with rotational flow. Because of this, clustering conditions derived from these equations, such as clustering of slow, Alfv\'en, and Suydam modes \citepads{2004JPlPh..70..651W} and clustering of thermal modes considered in this work, are altered by azimuthal flow. The expressions for the continua were studied in more detail. They are only altered by flow due to the additional Doppler shift to the frequency.

The analytic expressions were verified using the eigenvalue code \textit{Legolas}. The force-free helical magnetic field based on \citetads{1960MNRAS.120...89G}, helical flow field and coronal parameters mimic the environment to form a solar tornado. The true nature of solar tornadoes is a contested topic \citepads{2023SSRv..219...33G}. It should be noted that we are interested in thermal instability with background flow and only use the tornadoes as examples of condensations formed in the solar corona. The magnetic field of the benchmark case was predominantly vertical, with a smaller azimuthal component. It was set up this way because structures in the solar corona, such as coronal loops and prominences feet, have been observed or set up in numerical simulations vertically. All analytic expressions match the numerical results. For the benchmark case the thermal continuum is unstable. Slightly damped discrete Alfv\'en modes accumulate to a maximum in the Doppler shifted Alfv\'en continuum solely arising due to the background flow. The choice of cooling curve can cause the thermal continuum to take a different shape in the spectral plane. The location of the most unstable modes can vary from the inside to the outside. This has significant implications. The most unstable modes govern the behaviour of the plasma as a whole. Since thermal continuum modes are hyper-localised, the location can thus be altered just by chosing a different cooling curve. Thermal conduction damps out small perturbations. This has been known in literature \citepads{1965ApJ...142..531F} and has been shown in this work.

We derived an approximate expression to check whether discrete thermal and slow modes cluster towards an internal extremum of their respective continuum in the manner of \citetads{1991SoPh..134..247V}. \cref{Eq: omegaapprox} is not a closed criterion due to the unknown local wavenumber $q$ that it contains. However, using the sign of $\zeta$, given by \cref{Eq: zeta} and obtained through the WKB analysis, it can still be used to predict the appearance of the discrete modes. Using \textit{Legolas} we verified the expression for three equilibria. Two of them have discrete thermal modes above the continuum. For these the formation process of the condensations is governed by the eigenfunctions of the discrete modes. We therefore visualised the eigenfunctions in 2D spatial and temporal slices and in 3D. The distribution of the condensation due to the density perturbation has been shown. It is not a uniform condensation or singular condensation, due to the complicated shape of the eigenfunction and the dependence of the assumed waveform on the spatial coordinates and wavenumbers. 3D visualisation of the total velocity field shows that the field retains its helical shape but is also heavily influenced by the radial velocity perturbation. The choice of $m=1$, is clearly visible in the 2D and 3D plots. It would be interesting to investigate whether there are discrete thermal modes for different wavenumbers and how their eigenfunctions behave in space and time. \citetads{2023MNRAS.518.6355S} investigated the behaviour of adiabatic $m = 0$ and $m = \pm 1$ modes in a flux tube with rotational flow and showed that the $m=0$ modes are not influenced that much by the flow. The flux tube that they considered is embedded in an external photosphere. Just as in the well-known work by \citetads{1983SoPh...88..179E}, this kind of setup leads to surface and body modes, depending on where they propagate. We, however, did not consider an external medium as we are mainly focused on the thermal modes, which are internal due to their local nature.

Multidimensional MHD simulations are an interesting way to further study the formation of condensations in helical configurations with flow. Due to the one-dimensional, radial, variation of the equilibria used in MHD spectroscopy, physical effects depending on other coordinates are neglected. In the context of prominences and solar tornadoes gravity and the support against it are of the utmost importance. In MHD simulations this can be included and parametric studies can be performed to see what kind of flows and magnetic field configurations are needed to support the prominence mass in helical field, futher extending the analytic work by \citetads{2015ApJ...808L..23L}. The formation and appearance of condensations might be more transient. Comparing synthetic images with observations can teach us a lot about the existence of solar tornadoes. A second important physical improvement is the inclusion of a realistic solar atmosphere with density and temperature stratification, as recently used in e.g. \citetads{2021A&A...646A.134J} and \citetads{2023A&A...670A..64J}. Different prescriptions for background heating can be used. Models varying by height alone or as a function of local density and magnetic field are commonly used. They have been shown to affect the location and morphology of prominence formation \citepads{2022A&A...668A..47B}. The interplay between fragmentation of the condensations in the far non-linear stage, as shown by \citetads{2021A&A...655A..36H}, and flow is definitely worth investigating.

It must be stressed that the analytic equations derived in this work are valid for more physical environments than solely the solar corona. The solar coronal setup, mimicking possible tornado formation, was just used as an example. Thermal instability is a general phenomenon based on energy loss by radiation and flows are present on every scale in the universe. Therefore, there are many more interesting applications, such as the clumpiness of galactic ouflows \citepads{2020A&ARv..28....2V} and three-phase nature of the dynamic interstellar medium \citepads{2005ARA&A..43..337C}.

\begin{acknowledgements}

We wish to thank the anonymous referee, for the constructive comments that improved the paper, and the editorial office. JH would like to thank N. Claes and J. De Jonghe for the \textit{Legolas} support and V. Jer{\v{c}}i{\'c} and N. Brughmans for the useful discussions. The visualisations were achieved using the open source software Python (\url{http://python.org}) and ParaView (\url{http://paraview.org}). Both authors are supported by the ERC Advanced Grant PROMINENT and a joint FWO-NSFC grant G0E9619N. This project has received funding from the European Research Council (ERC) under the European Union’s Horizon 2020 research and innovation programme (grant agreement No. 833251 PROMINENT ERC-ADG 2018). This research is further supported by Internal funds KU Leuven, project C14/19/089 TRACESpace.

\end{acknowledgements}

\bibliographystyle{aa}
\bibliography{main.bib}

\end{document}